\documentclass[]{spie}  

\usepackage{amsmath,amsfonts,amssymb}
\usepackage{graphicx}
\usepackage[colorlinks=true, allcolors=blue]{hyperref}

\usepackage{times}
\usepackage{float}
\usepackage{booktabs}
\usepackage{epstopdf}
\usepackage[table]{xcolor}
\usepackage{amsmath}
\usepackage{numprint}
\usepackage{subcaption}
\usepackage{caption}
\usepackage{multirow}
\usepackage{array}
\usepackage{calc}
\usepackage{calc}
\usepackage{tikz}
\usetikzlibrary{positioning}
\usepackage{algorithmic}
\usepackage{algorithm}

\title{Deep Learning-Based Defect Classification and Detection in SEM Images}

\author[a,c]{Bappaditya Dey}
\author[f]{Dipam Goswami}
\author[a]{Sandip Halder}
\author[b,d]{Kasem Khalil}
\author[a]{Philippe Leray}
\author[c,e]{Magdy A. Bayoumi}
\affil[a]{imec, Kapeldreef 75, 3001 Leuven, Belgium}
\affil[b]{}
\affil[c]{The Center for Advanced Computer Studies,  University of Louisiana at Lafayette, Louisiana, USA}
\affil[d]{Department of Electrical Engineering, Assiut University, Egypt}
\affil[e] {Department of Electrical and Computer Engineering, University of Louisiana at Lafayette, Louisiana, USA}
\affil[f]{Department of Mathematics, Birla Institute of Technology and Science, Pilani, India}

\authorinfo{Further author information: (Send correspondence to Bappaditya Dey and Kasem Khalil)\\Bappaditya Dey: E-mail: Bappaditya.Dey@imec.be}

\begin{document} 
\maketitle

\begin{abstract}
	Defect inspection in semiconductor processes has become a challenging task due to continuous shrink of device patterns (pitches less than 32 nm) as we move from node to node. Current state-of-the-art defect detection tools (optical/e-beam) have certain limitations as these tools are driven by some rule-based techniques for defect classification and detection. These limitations often lead to misclassification of defects, which leads to increased engineering time to correctly classify different defect patterns. In this paper, we propose a novel ensemble deep learning-based model to accurately classify, detect and localize different defect categories for aggressive pitches and thin resists (High NA applications).In particular, we train RetinaNet models using different ResNet, VGGNet architectures as backbone and present the comparison between the accuracies of these models and their performance analysis on SEM images with different types of defect patterns such as bridge, break and line collapses. Finally, we propose a preference-based ensemble strategy to combine the output predictions from different models in order to achieve better performance on classification and detection of defects. As CD-SEM images inherently contain a significant level of noise, detailed feature information is often shadowed by noise. For certain resist profiles, the challenge is also to differentiate between a microbridge, footing, break, and zones of probable breaks. Therefore, we have applied an unsupervised machine learning model to denoise the SEM images to remove the False-Positive defects and optimize the effect of stochastic noise on structured pixels for better metrology and enhanced defect inspection. We repeated the defect inspection step with the same trained model and performed a comparative analysis for “robustness” and “accuracy” metric with conventional approach for both noisy/denoised image pair. The proposed ensemble method demonstrates improvement of the average precision metric (mAP) of the most difficult defect classes. In this work we have developed a novel robust supervised deep learning training scheme to accurately classify as well as localize different defect types in SEM images with high degree of accuracy. Our proposed approach demonstrates its effectiveness both quantitatively and qualitatively.\\
\end{abstract}



\section{Introduction}
\label{sect:intro}  
As we scale from node-to-node, device dimensions become smaller and smaller, and this brings unprecedented challenges to optical inspection as well as for e-beam inspection. Recently, e-beam based inspection has become more and more pertinent for extremely small defect detections. The graph below shows the inspection space w.r.t the available tools. For inspection e-beam is often more sensitive when compared to optical but classification remains a challenge for both methods. Also, defect location accuracy is better for e-beam based tools which are often linked to design databases. Even though resolution and location accuracy improve greatly with e-beam tools, absence of a robust classification algorithms often lead to increased engineering time (as engineers manually classify defects). We notice that even on commercially available software, classification is not robust. This has forced us to look for alternative methods/algorithms for more robust defect classification.

\begin{figure}
	\begin{center}
		\begin{tabular}{c}
			\includegraphics[height=7.5cm]{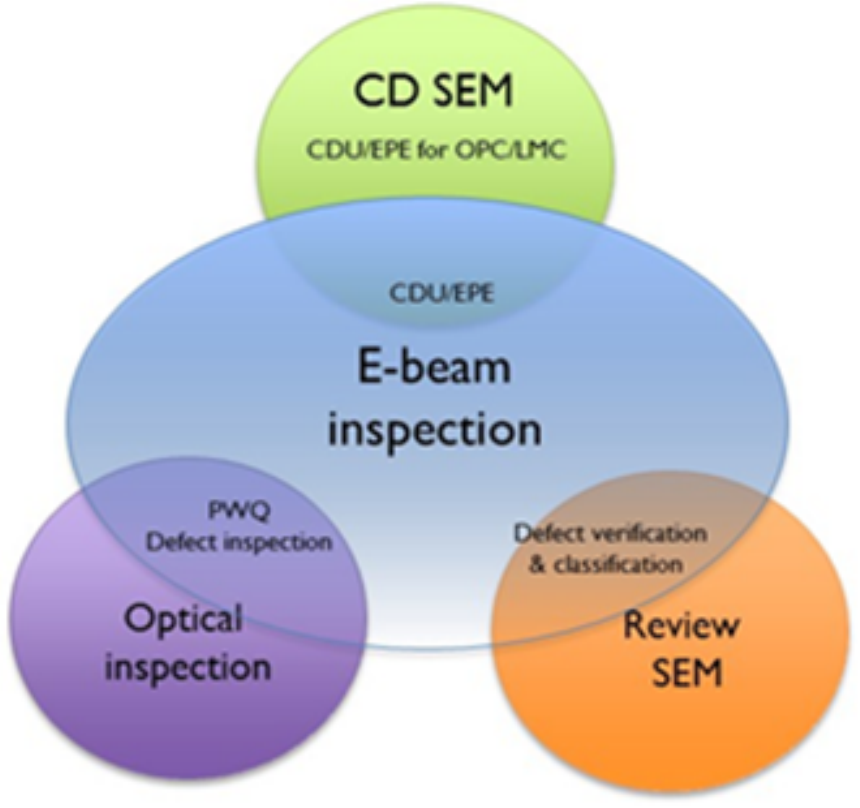}
		\end{tabular}
	\end{center}
	\caption 
	{ \label{fig:Defect}
		The Defect Inspection and Review Space. } 
\end{figure}

Fig.~\ref{fig:Defect} Shows the defect inspection and review space. E-beam inspection tools cover a multitude of different applications. While there have been some publications in the past in semiconductor manufacturing for BEOL applications Ref.~\citenum{8373144}, there are not many for EUV defects, especially for bridges and collapses which are not trivial and where many commercial software often fail. 

These defects may range from being critical failures to wafer-yield limiters. Fig.~\ref{fig:Defects2} shows SEM images with examples of different defect categories generally encountered in aggressive pitches. Fig.~\ref{fig:Defects2}(a) , (b) and (c) are examples of Line-Space (L/S) patterns with defect type Bridge, Line-Collapse and Broken-line/Gap category, respectively. We depicted more challeging defect scenarios in Fig.~\ref{fig:Defects2}(d), as an example of Broken-line/Gap and probable-gap (partial feature missing) defect types in presence of contrast/intensity change (image intensity sometimes vary strongly from one line in the SEM image to another line due to different charging, when a line is broken, somewhere, not necessarily in the FOV) as well as in Fig.~\ref{fig:Defects2}(e), presence of random micro-bridges with variable degrees of pixel-level defect. The goal of this work is to show how Deep Learning (DL)-based algorithms can be used for more robust classification of different defects during wafer processing after an optimal focus/dose is selected. In a previous work Ref.~\citenum{Bap} we have already shown the benefit of using such techniques to help in drawing process windows automatically from FEM wafers. In this paper, we go a step further and show how such DL methods can classify the tougher bridge/collapse defects together with other process defects. In summary, there are four contributions in this work:

\begin{figure}
	\begin{center}
		\begin{tabular}{c}
			\includegraphics[width=.700\linewidth]{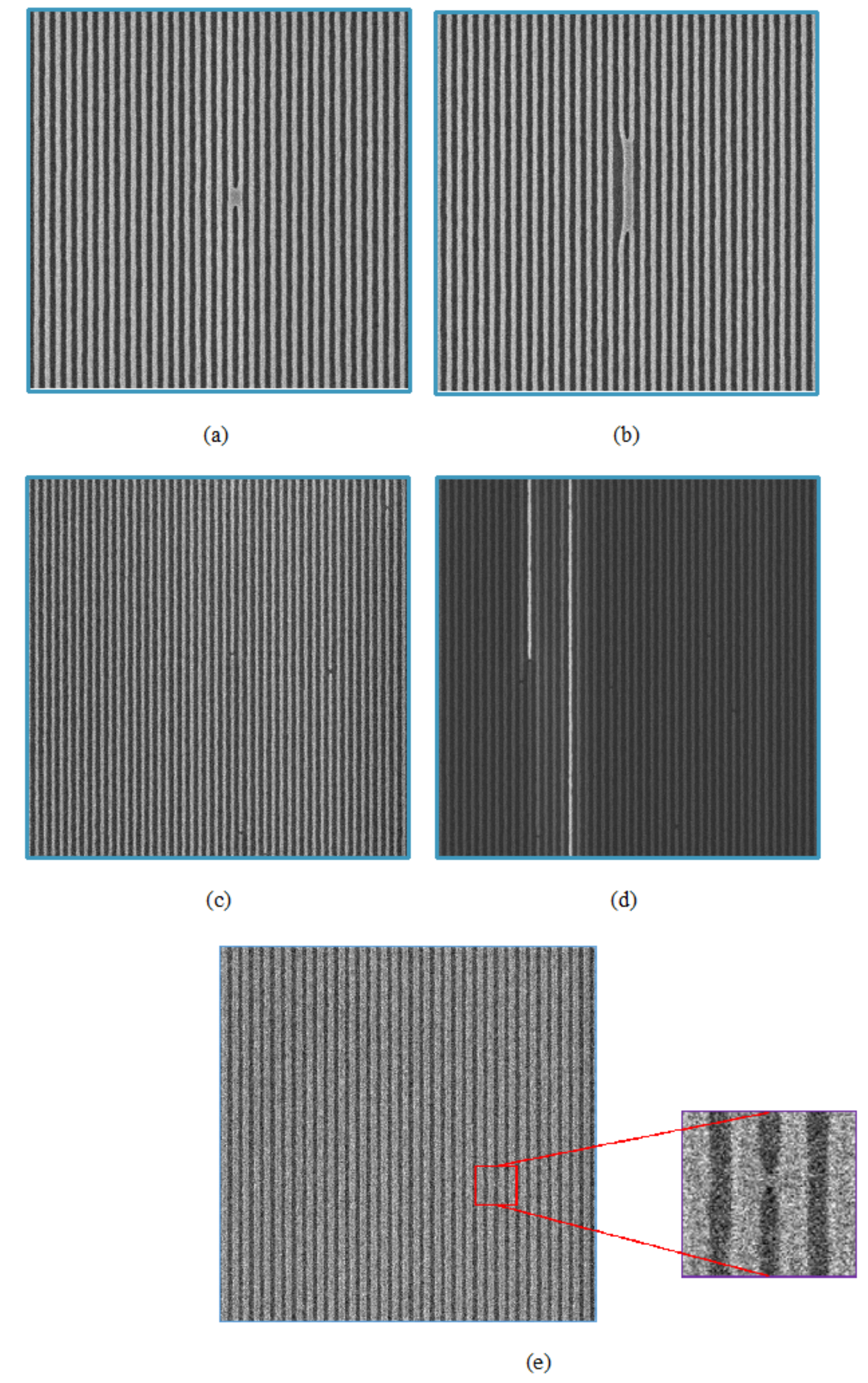}
		\end{tabular}
	\end{center}
	\caption 
	{ \label{fig:Defects2}
		Typical Defects: (a) Bridge, (b) Line-Collapse, (c) Gap and Prob-Gap, and 
		Challenging Defects: (d) Gap and Prob-Gap in presence of contrast change (e) Micro-bridges. } 
\end{figure} 
\begin{itemize} 
	\item A novel ensemble deep-learning model is proposed to solve challenging defect detection problems in SEM images. Our goal is to accurately classify, detect and localize different defect categories for aggressive pitches and thin resists (High NA applications). We have trained RetinaNet models using different ResNet Ref.~\citenum{he2015deep}, VGGNet Ref.~\citenum{DBLP:journals/corr/SimonyanZ14a}, SSD$\_$MobileNet$\_$v1 Refs.~\citenum{Hinton504, 10.1007/978-3-319-46448-0_2} architectures as backbone and proposed a preference-based ensemble strategy to combine the output predictions from different models and achieve better performance on classification and detection of different defects. The proposed ensemble method demonstrates improvement of the average precision metric (mAP) of the most difficult defect classes.
	\item Defect inspection in ADI SEM (After Develop Inspection) images is the most challenging task as different sources of noise  Ref.~\citenum{timischl2012statistical} generally shadow the detailed device feature information. This often leads to false defect detections and erroneous metrology. The challenge also lies for some resist profile, in diffrentiation and detection of minute bridges (micro), breaks (zones of probable breaks) and resist footing from these noisy SEM images. Therefore, we have applied an unsupervised machine learning strategy to denoise Refs.~\citenum{Bap2} the SEM images aiming to optimize the effect of stochastic noise on structured pixels and therefore, to remove the False-Positive defects (FP) for better metrology and enhanced defect inspection. We have repeated the defect inspection step with the same trained model parameters and have performed a comparative analysis for “robustness” and “accuracy” metric for both noisy/denoised image pairs with different “detection confidence score”. We also have fine-tuned the proposed model by training with denoised images for the above-mentioned challenging defect classes.

	\item We have analysed and validated our proposed model performance against conventional tools or approaches. We have noticed that while using the conventional approach, various defects are not being flagged and we believe that this limitation is due to the “manual” selection of the detection threshold parameter. Furthermore, the detection scenario is influenced by the condition if the image is noisy or denoised. However, our proposed model demonstrates “stable” performance in detecting defects with better accuracy for both noisy or denoised images and replaces the manual trial-and-error based “threshold” selection method with automated “confidence score”. Once defects are correctly detected, different parameters (as length, width, area, additional feature vectors) about the defects can be output for better understanding the root cause of the defects. Thus, Our proposed approach demonstrates its effectiveness both quantitatively and qualitatively. 
	
	\item Finally, we built an UI (User-Interface) using Streamlit library Ref.~\citenum{17} to deploy our proposed model as a web-based defect inspection app. This will enable different partners/vendors to run the application on their local servers/workstations on their own tool data. This UI will enable the users to upload a dataset of SEM/EDR/Review-SEM images, to select and run the ineference model on the dataset, to visualize the prediction performance locally and finally to segregate and save the images in different folders according to their defect categorical classes in local machines.
\end{itemize}

The remainder of the paper is organized as follows. Sec. 2 introduces some related work. In Sec. 3, we provide an overview of the RetinaNet architecture backbone and our proposed ensemble method. Sec. 4 demonstrates the experiments done followed by Sec. 5 covering the performance evaluation and comparison analysis. In Sec. 6, we conclude the paper.

\section{Related Work} 
\label{sect:Related Work}
In this section, we briefly discuss existing research approaches and methodologies in the context of machine learning based defect inspection. Our search criteria are only limited to semiconductor process domain. Convolutional Neural Network Ref.~\citenum{khalil2022designing} or simply ConvNet gained popularity in the domain of computer vision applications following Yann LeCun’s first introduced LeNet architecture Ref.~\citenum{726791}, aimed to recognize handwritten digits. Since then, researchers have experimented with more complex architecture variants of recurent neural network Ref.~\citenum{khalil2019economic} and CNN, as AlexNet Ref.~\citenum{10.5555/2999134.2999257}, GoogleNet Ref.~\citenum{7298594}, VGGNet Ref.~\citenum{DBLP:journals/corr/SimonyanZ14a}, ResNet Ref.~\citenum{he2015deep}, RetinaNet Ref.~\citenum{8237586} etc., to correlate depth of architectures, especially convolutional blocks with model accuracy. Recurrent usage of these deep learning models can also be observed for robust defect inspection strategy in every single process step of any real-world production or manufacturing pipeline. Semiconductor industries are of no exception to this. M. Sharifzadeh \textit{et. al.} Ref.~\citenum{4777721} has investigated detection and classification of four steel surface defect categories as hole, scratch, coil break and rust using conventional image processing algorithms. Authors reported four different accuracy metrics for detecting above four popular kind of steel defects. The drawback of this proposed method is the “trial and error selection method” of the high-performance method among several tested image processing algorithms. Four most common operations for defect detection as (1) thresholding, (2) noise removal, (3) edge detection and (4) segmentation do not survive the requirements of the semiconductor industry at this advanced node era. S. Kim \textit{et. al.} Ref.~\citenum{Kim2017AutomaticDD} proposed a novel defect detection approach using component tree representations of SEM images. They proposed a modified version of the original framework to better identify defects. The algorithm is based on the topographic map representation of the SEM images and therefore built a component tree in quasi-linear time with respect to attributes such as area, height, volume, and stiffness etc. These attributes proved to be essential to define and detect meaningful defect regions. The authors proposed two versions of the framework, one to detect defects of uniform size and shape and another to tackle more difficult cases with variable size and shape. The drawback of this methodology is that the performance evaluation is only qualitative and not quantitative. There exists no standard dataset or standard evaluation protocol to carry out fair comparative analysis of this proposed approach. Ravi Bonam \textit{et. al.} Ref.~\citenum{6552751} has studied the effect of defect sizes and their impact on EUV lithography. It is observed that optimal value of manually tuned process parameters depends on defect types when technology node scale is of few nanometers. Also, the capture efficiency of the manual technique is directly proportional to the nuisance rate. J. Wang \textit{et. al.} Ref.~\citenum{8765895} proposed “AdaBalGAN” model (adaptive balancing generative adversarial network) to tackle misidentification problem of defect pattern recognition from wafer maps due to imbalanced defective class data. An adaptive generative model is proposed to balance the number of samples of each defect category as per classification accuracy. Ji-Hee Lee \textit{et. al.} Ref.~\citenum{Lee_2019} has proposed transfer learning mechanism in the scope of machine learning strategy to build a reliable defect detection method for patterned wafer images. Their idea is based on to cope with new categories of defect data, with an already trained model, in each update cycle with minimum data possible and with minimum engineering resources and time. They have shown how deep learning method is outperforming than traditional defect inspection algorithms both in accuracy and time. The drawback of this strategy is that it can only judge whether a defect exists or not but not able to classify the defect classes. B. Devika \textit{et. al.} Ref.~\citenum{8944584} proposed a CNN based deep learning model to identify wafer defect patterns. The authors have taken into consideration the Wafer Bin Map (WBM) patterns as circle, cluster, scratch, and spot since each defect pattern correlates with different fabrication errors. An early and efficient ML based defect detection strategy leads to reduced wafer test time and an improved die yield. Jong-Chih Chien \textit{et. al.} Ref.~\citenum{app10155340} has proposed two ways to use deep CNN architecture to classify semiconductor defect images. The defect classes were aimed as center, local, random and scrape. Their approaches were not tested on mixed defect types as well as the authors reported about occurrence of misclassification during validation phase which demands further fine-tuning of the model. Y. Yuan-Fu Ref.~\citenum{8791815} proposed alternative machine learning techniques against automatic optical inspection (AOI) to visualize defect patterns and to identify root causes of die failures. The limitation of AOI method is that it still requires human expert intervention to judge the type of defect. In this paper, the authors proposed CNN and XGBoost techniques to retrieve wafer maps and to classify the defect patterns. They compared the classification performance of the proposed method against random decision forests (RF), support vector machine (SVM), adaptive boosting (Adaboost). Dhruv V. Patel \textit{et. al}. Ref.~\citenum{10.1117/1.JMM.19.2.024801} demonstrated the effectiveness of optimized deep learning models in identifying, localizing, and classifying different types of wafer defects with high degree of accuracy. They have generated high-resolution EB images of wafers patterned with different types of intentional defect categories and trained their CNN based models. They have also demonstrated the significance of CAM (Class Activation Maps) for localizing the defects. Joongsoo Kim \textit{et. al.} Ref.~\citenum{8373144} proposed a CNN based defect image classification model derived from Residual Network to classify defects specialized for TSV (Through-Silicon-Via) process. Image preprocessing has been performed before the model deployment to increase classification accuracy as well as to tackle size dependent defect classification issue.

We have carefully examined all the proposed methodologies, solutions addressed and most importantly the limitations as discussed by previous authors to formulate our proposed approach.

\section{Proposed Approach}
\label{section:Proposed Approach}
\begin{figure*}
	\begin{center}
		\begin{tabular}{c}
			\includegraphics[width=.800\linewidth]{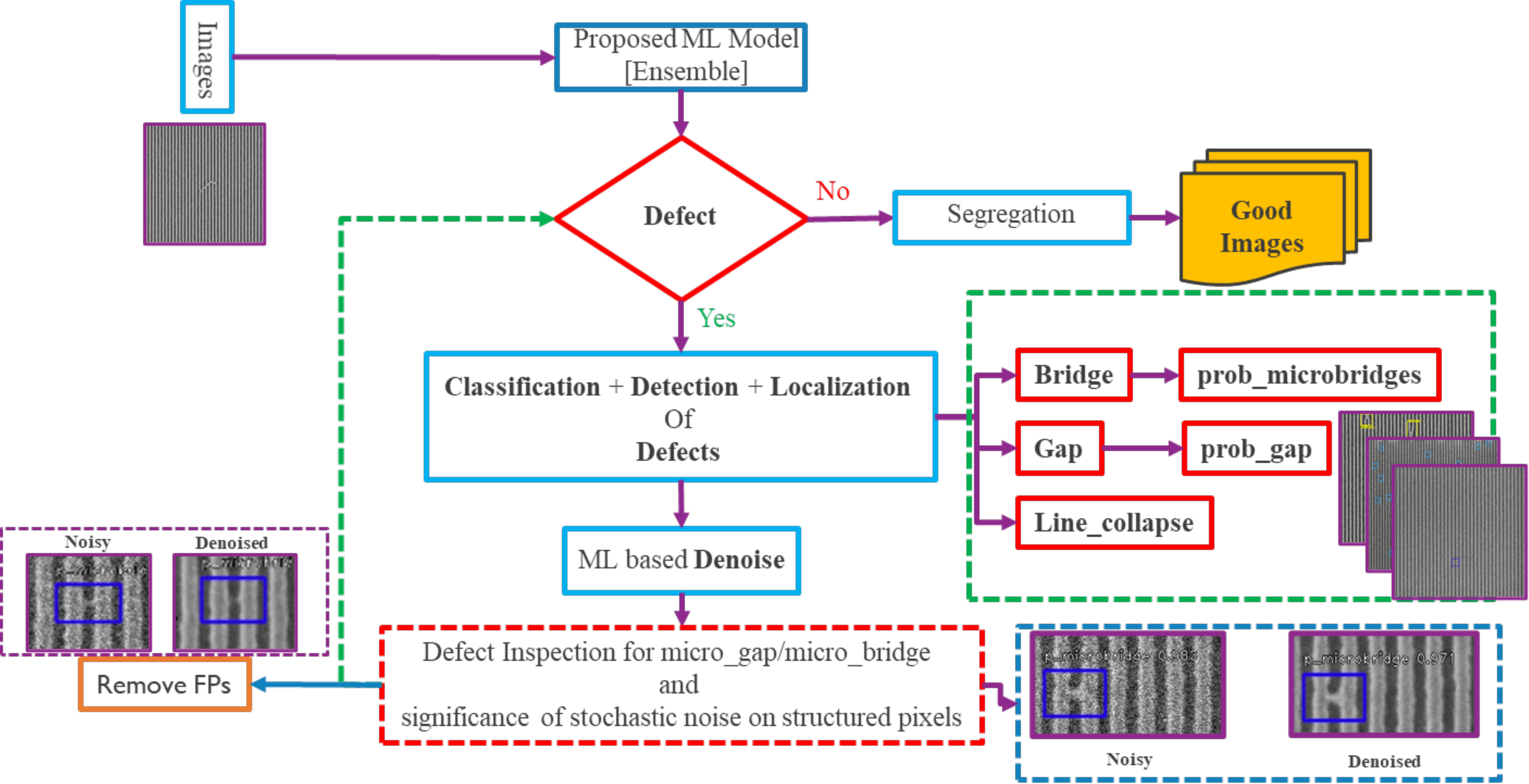}
		\end{tabular}
	\end{center}
	\caption 
	{ \label{fig:Proposed}
		Proposed ensemble model-based defect detection framework. } 
\end{figure*}

In this section, our proposed approach, based on RetinaNet, to detect different defects from SEM images and classify them according to their corresponding classes in aggressive pitches is presented. Our proposed ensemble model-based defect detection framework, which is illustrated in Fig.~\ref{fig:Proposed}, consists of the RetinaNet based detector and the U-Net architecture based denoiser. The framework is trained and evaluated using imec datasets (both Post-Litho and Post-Etch resist Wafer dataset) and classifies, detects, and localizes the candidate defect types. The focus of this section is to briefly discuss the key modules of the defect detection network only. We have utilized U-Net architecture based deep learning denoiser following Refs.~\citenum{Bap2, dey2021unsupervised} and therefore beyond the scope of this research. To the best of our knowledge, this framework is the first to apply a novel robust supervised deep learning training scheme to accurately classify as well as localize different defect types in SEM images. The key modules of the defect detection network are:

\begin{itemize} 
	\item RetinaNet defect detector architecture
	\item Deep feature extractor networks as backbone
\end{itemize}

\subsection{Overview of the RetinaNet architecture}
\begin{figure*}
	\begin{center}
		\begin{tabular}{c}
			\includegraphics[width=.800\linewidth]{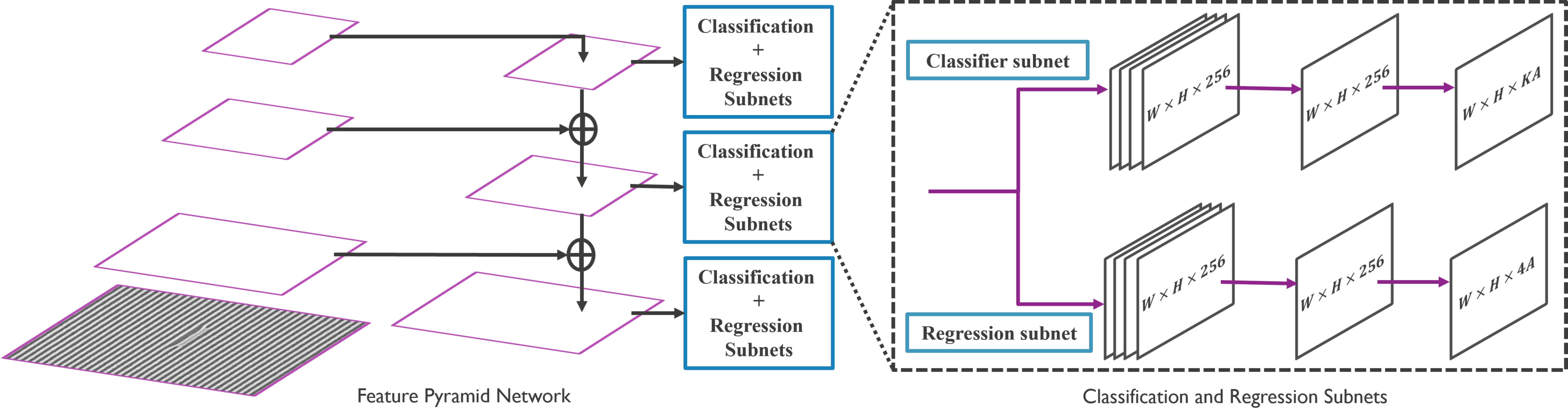}
		\end{tabular}
	\end{center}
	\caption 
	{ \label{fig:RetinaNet}
		RetinaNet defect detector architecture. } 
\end{figure*}

RetinaNet is a popular one-stage object detection model which works well with dense objects and effectively handles the foreground-background class imbalance problem affecting the performance of other one-stage detector models. RetinaNet architecture consists of a Feature Pyramid Network (FPN) Ref.~\citenum{lin2017feature} built on top of a deep feature extractor network, followed by two subnetworks, one for object classification and the other for bounding box regression. RetinaNet defect detector architecture is illustrated in Fig.~\ref{fig:RetinaNet}.

FPN takes one single resolution input image, subsamples it into multiple lower resolution images and outputs the feature maps at different scales, thus building a multi-scale feature pyramid representation. Thus, it enables detection of objects of varying sizes from different layers of the feature pyramid. FPN combines low resolution features with high resolution features via a top-down pathway which has lateral connections to layers from a bottom-up pathway. The bottom-up pathway generates a feature hierarchy using feature maps of different scales from the input image. The top-down pathway performs up-sampling on the spatially coarser feature maps coming from the higher pyramid levels. The lateral connections are then used to merge the feature maps of same spatial size from both the paths which gives semantically strong feature maps.

The classification and regression subnetworks are connected to every layer of the feature pyramid and are independent from each other. The classification subnetwork predicts the probability for the presence of an object for every anchor box and object class. It consists of 4 fully convolutional layers [(3×3) conv layer with 256 filters and ReLU activation]. It follows another (3×3)convolutional layer having K×A filters where K is the number of classes and A is the number of anchors (A=9 anchors covering 3 different aspect ratios and 3 different scales). 

The regression subnetwork is used for regressing the offset for the anchor boxes against the ground truth object boxes. It is a class-agnostic regressor which does not know what class the objects belong to and uses fewer parameters. The structure is similar to the classification network except that it outputs 4 bounding box coordinates for every anchor box. Anchor boxes are assigned to ground truth boxes if the IOU between the boxes exceeds 0.5 and assigned to background if the IOU is in the range [0,0.4). The anchor boxes having IOU in the range [0.4,0.5) are ignored.

RetinaNet uses focal loss Ref.~\citenum{8237586} which improves the prediction accuracy giving more importance to the hard samples during training and reducing the contribution of easy samples to the loss. It enhances the cross-entropy loss by introducing a weighting factor to offset the impact of class imbalance and a modulating factor to focus more on training the hard negatives and less on the easy examples.

\subsection{Deep feature extractor networks as backbone}
\begin{figure*}
	\begin{center}
		\begin{tabular}{c}
			\includegraphics[width=0.800\linewidth]{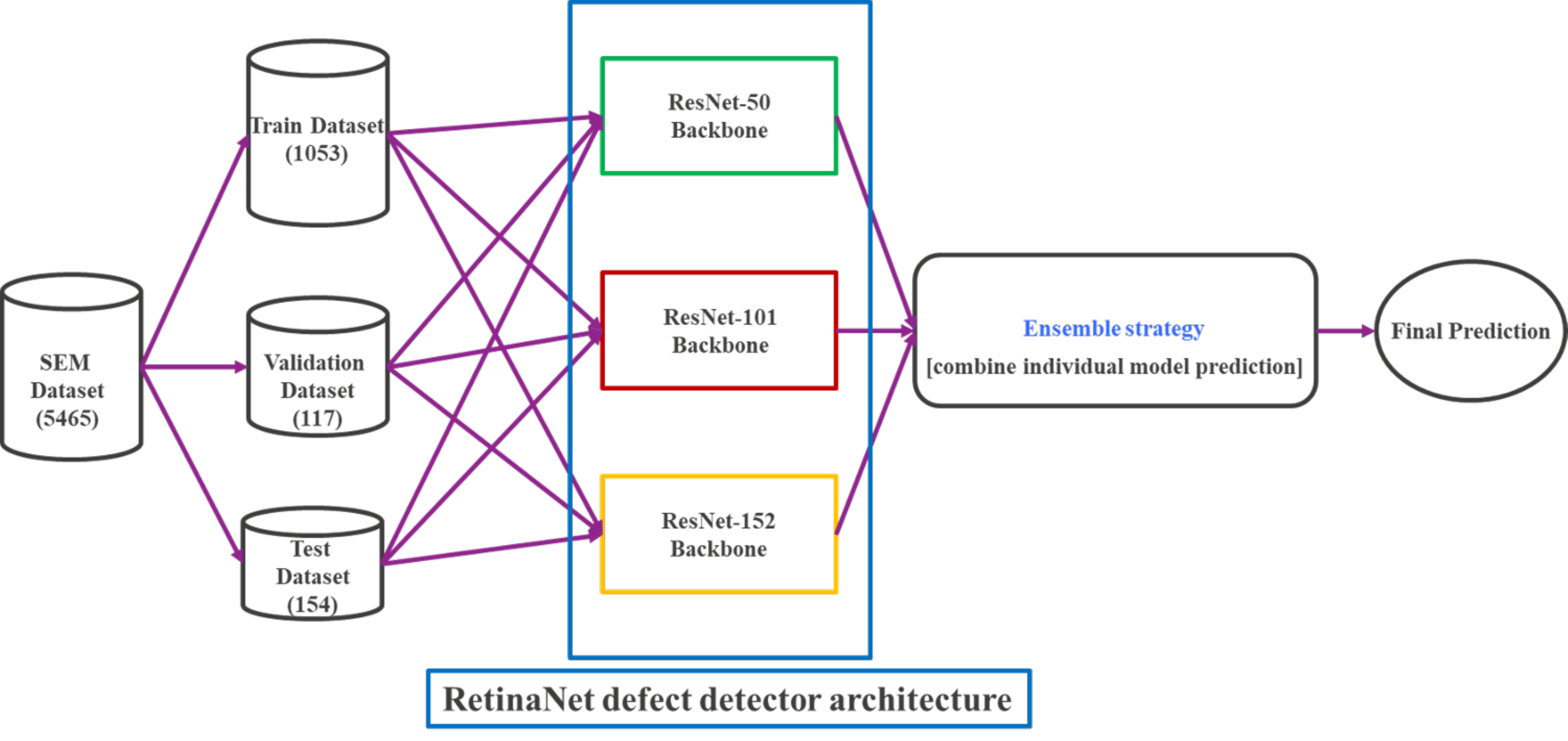}
		\end{tabular}
	\end{center}
	\caption 
	{ \label{fig:Deep}
		Deep feature extractor networks as backbone. } 
\end{figure*}

The proposed RetinaNet defect detector framework is an ensemble architecture based on a selective permutation of backbones as ResNet50, ResNet101 and ResNet152 as shown in Fig.~\ref{fig:Deep}. Table~\ref{tab:RESNET50} describes custom variants of ResNet architectures with multiple convolutional layers with skip connections across them for feature extraction and fully connected layers for predicting different defect category probabilities. We have taken the affirmative ensemble Ref.~\citenum{CasadoGarcia19} of the predictions from the 3 ResNet models with preference to the models showing better performance on the test dataset. So, we consider all the predictions from the first model and then we add those predictions from the second-best model which are not overlapping with the first model predictions. We use an IOU threshold of 0.5 to consider the boxes as overlapping. In this way, we add the non-overlapping predictions from the third-best model. This ensemble strategy ensures that all the predictions from the 3 models are taken, and this improves the accuracy of the test dataset.

\begin{table*}[ht]
	\caption{RESNET50, RESNET101 AND RESNET152 BACKBONE ARCHITECTURE.} 
	\label{tab:RESNET50}
	\begin{center}       
		\begin{tabular}{|l|l|l|l|} 
			\hline

			\rule[-1ex]{0pt}{3.5ex} Layer Name	& ResNet-50 Backbone	 & ResNet-101 Backbone	& ResNet-152 Backbone \\
			\hline
			
			\rule[-1ex]{0pt}{3.5ex} Conv1 & \multicolumn{3}{c|}{7×7, 64, stride 2} \\
			\hline
			
			\rule[-1ex]{0pt}{3.5ex} \multirow{2}{*}{} &\multicolumn{3}{c|}{3×3, max pool, stride 2}\\\cline{2-3}
			\hline
			
			\rule[-1ex]{0pt}{7.5ex} Conv2$\_$x & $ \begin{bmatrix}  1 \times 1& ,  & 64\\
			3 \times 3 & ,  & 64\\
			1 \times 1 & ,  & 256 \end{bmatrix}$ $\times 3$ &
			$\begin{bmatrix}
			1 \times 1& ,  & 64\\
			3 \times 3 & ,  & 64\\
			1 \times 1 & ,  & 256
			\end{bmatrix}$ $\times 3$& 
			$	\begin{bmatrix}
			1 \times 1& ,  & 64\\
			3 \times 3 & ,  & 64\\
			1 \times 1 & ,  & 256
			\end{bmatrix}$ $\times$ 3\\\cline{2-3}
			\hline
			
			\rule[-1ex]{0pt}{7.5ex} Conv3$\_$x& 
			$\begin{bmatrix}
			1 \times 1& ,  & 128\\
			3 \times 3 & ,  & 128\\
			1 \times 1 & ,  & 512
			\end{bmatrix}$
			$\times 4$& $
			\begin{bmatrix}
			1 \times 1& ,  & 128\\
			3 \times 3 & ,  & 128\\
			1 \times 1 & ,  & 512
			\end{bmatrix}
			$ $\times 4$& $
			\begin{bmatrix}
			1 \times 1& ,  & 128\\
			3 \times 3 & ,  & 128\\
			1 \times 1 & ,  & 512
			\end{bmatrix}
			$ $\times 8$\\
			\hline
			
			\rule[-1ex]{0pt}{7.5ex} Conv4$\_$x& $
			\begin{bmatrix}
			1 \times 1& ,  & 256\\
			3 \times 3 & ,  & 256\\
			1 \times 1 & ,  & 1024
			\end{bmatrix}
			$ $\times 6$& $
			\begin{bmatrix}
			1 \times 1& ,  & 256\\
			3 \times 3 & ,  & 256\\
			1 \times 1 & ,  & 1024
			\end{bmatrix}
			$ $\times 23$ & $
			\begin{bmatrix}
			1 \times 1& ,  & 256\\
			3 \times 3 & ,  & 256\\
			1 \times 1 & ,  & 1024
			\end{bmatrix}
			$ $\times 36$\\
			\hline

			\rule[-1ex]{0pt}{7.5ex} Conv5$\_$x& $
			\begin{bmatrix}
			1 \times 1& ,  & 512\\
			3 \times 3 & ,  & 512\\
			1 \times 1 & ,  & 2048
			\end{bmatrix}
			$ $\times 3 $& $
			\begin{bmatrix}
			1 \times 1& ,  & 512\\
			3 \times 3 & ,  & 512\\
			1 \times 1 & ,  & 2048
			\end{bmatrix}
			$ $\times 3 $& $
			\begin{bmatrix}
			1 \times 1& ,  & 512\\
			3 \times 3 & ,  & 512\\
			1 \times 1 & ,  & 2048
			\end{bmatrix}
			$ $\times 3$\\
			\hline
			
			\rule[-1ex]{0pt}{3.5ex} FC& \multicolumn{3}{c|}{Average pool, 1000-d fc, softmax}\\
			\hline

		\end{tabular}
	\end{center}
\end{table*}

The goal of our proposed ensemble model-based defect detection framework is based on 2 major steps. In the first step, train a RetinaNet defect detector architecture as discussed above to accurately classify as well as localize different defect types in SEM images such as bridge, line$\_$collapse, gap, micro-bridges, and micro-gaps, respectively. In the second step, denoise those SEM images mainly with challenging defects such as micro-bridges and micro-gaps to optimize the effect of stochastic noise on structured pixels and reiterate the defect detection step to remove the False-Positive defects (FP) towards better metrology and enhanced defect inspection.

\section{Experiments}
\label{section: Experiments}
Our proposed ensemble model-based defect detection framework is implemented using Keras library Ref.~\citenum{chollet2015keras} and the Tensorflow library Ref.~\citenum{tensorflow2015-whitepaper} backend in the python programming environment. The Anaconda version was 4.9.2. Our model has been trained and evaluated on Lambda TensorBook with NVIDIA RTX 2080 MAX-Q GPU.

\subsection{Datasets}
The proposed ensemble model (Classifier + Detector) is trained and evaluated on both Post-Litho and Post-Etch P32 (Pitch 32 nm) resist Wafer dataset. The dataset consists of a total of 5,465 raw SEM images of $(1024 \times 1024)$ pixels in TIFF format with stochastic defects such as bridge, line-collapse, gaps/line-breaks, micro/nano-bridges, and probable nano-gaps as well as clean images without any such defects. The representative defect class images from this dataset are already shown in Fig.~\ref{fig:Defects2} (a) – (e). We have manually labeled 1324 SEM images (1053 training images, 117 validation images, and 154 test images) images using LabelImg Ref.~\citenum{Ingo24} graphical image annotation tool. The defect labeling strategy comprises diverse defect representative and challenging condition instances and as per naming convention in Table~\ref{tab:CLASS}. The dataset is divided into a training set, a validation set, and a test set as shown in Table~\ref{tab:Data} . We have a total of 2529 defect instances of these 5 different defect classes for training and 337 instances for validation. To comply with training criteria, we converted all images with “.tiff” format into “.jpg” format. We also implemented different data-augmentation techniques (as rotation, translation, shearing, scaling, flipping along X-axis and Y-axis, contrast, brightness, hue, and saturation) to balance/increase the diversity of training dataset defect patterns. We did not consider using any digital twins or synthetic datasets as cited in some previous citations as that does not solve the purpose of tackling real FAB originated stochastic defectivity scenario. We have also excluded any fabricated dataset patterned with intentionally placed or programmed defect types.

\begin{table}[ht]
	\caption{DEFECT CLASS LABELING CONVENTION.} 
	\label{tab:CLASS}
	\begin{center}       
		\begin{tabular}{|l|l|} 
			\hline

			\rule[-1ex]{0pt}{3.5ex}\textbf{DEFECT CATEGORY}& \textbf{LABELLED AS}\\
			\hline
			
			\rule[-1ex]{0pt}{3.5ex} BRIDGE	&\textbf{bridge}\\
			\hline
			
			\rule[-1ex]{0pt}{3.5ex} LINE-COLLAPSE &\textbf{line$\_$collapse}\\
			\hline
			
			\rule[-1ex]{0pt}{3.5ex} GAP/LINE-BREAKS &\textbf{gap}  \\
			\hline
			
			\rule[-1ex]{0pt}{3.5ex} MICRO/NANO-BRIDGE &\textbf{microbridge} \\
			\hline
			
			\rule[-1ex]{0pt}{3.5ex} PROBABLE NANO-GAP &\textbf{p$\_$gap}  \\
			\hline

		\end{tabular}
	\end{center}
\end{table}


\begin{table}[ht]
	\caption{DATA DISTRIBUTION OF DEFECT SEM IMAGES.} 
	\label{tab:Data}
	\begin{center}       
		\begin{tabular}[6pt]{|m{2.5cm}|m{2.5cm}|m{2.5cm}| m{2.5cm}|}
			\hline

			\rule[-1ex]{0pt}{3.5ex} Class Name	& Train 	&Val  &	Test  \\

			\rule[-1ex]{0pt}{3.5ex} 	&  (1053 images)	& (117 images) &	(154 images) \\
			\hline
			
			\rule[-1ex]{0pt}{3.5ex} gap	&1046 &	156	&174 \\
			\hline
			
			\rule[-1ex]{0pt}{3.5ex} p$\_$gap	&315	&49	&54 \\
			\hline
			
			\rule[-1ex]{0pt}{3.5ex} microbridge	&380	&47	&78 \\
			\hline
			
			\rule[-1ex]{0pt}{3.5ex} bridge	&238&	19&	17  \\
			\hline
			
			\rule[-1ex]{0pt}{3.5ex} line$\_$collapse&	550&	66&	76  \\
			\hline
			
			\rule[-1ex]{0pt}{3.5ex} Total Instances	&2529	&337&	399  \\
			\hline

		\end{tabular}
	\end{center}
\end{table}

\subsection{Evaluation criteria}
We have considered Intersection over Union (IoU) Ref.~\citenum{8953982} between the ground truth bounding box and the predicted bounding box $\geq 0.5$. The “defect detection confidence score threshold” metric is taken as 0.5. The proposed ensemble model-based defect detector overall performance is evaluated against mAP as Mean Average Precision, where mAP is calculated using the weighted average of precisions among all defect classes. AP or average precision provides the detection precision for one specific defect class. We have also considered the speed of detection per image (average-inference-time in seconds/milliseconds). We have taken the affirmative ensemble Ref.~\citenum{CasadoGarcia19} of the predictions from top $k$ backbones with preference to the models showing better performance on the test dataset. So, we consider all the predictions from the first model and then we add those predictions from the second-best model which are not overlapping with the first model predictions. We use an IOU threshold of 0.5 to consider the boxes as overlapping. In this way, we add the non-overlapping predictions from the third-best model and so on up to k models. This ensemble strategy ensures that all the predictions from the top $k$ backbones are taken, and this improves the accuracy of the test dataset. The improvement is noticeable for the most difficult defect category p$\_$gap where the ensemble precision exceeds the individual model precisions.

\subsection{Training}
We have first trained RetinaNet model experimentally the different individual backbone architectures (ResNet50, ResNet101, ResNet152, SSD$\_$MobileNet$\_$v1, SeResNet34, Vgg19 and Vgg16) on our SEM image dataset as discussed in previous section independently. For the proposed experiments, we have selected training parameters and hyperparameters as: 40 epochs, batch-size of 1, initial learning rate at 0.00001, learning rate reduction by a factor of 0.1 if learning rate plateaus and optimizer as ADAM Ref.~\citenum{kingma2017adam}. Table~\ref{tab:IMPLEMENTATION} provides the comparison analysis for defect detection accuracies obtained per defect class as well as mAP on test images for the above experimental backbones with score-threshold 0.50, with AP50. Table~\ref{tab:OVERAL75} provides the comparison analysis for defect detection accuracies obtained per defect class as well as mAP on test images for the above experimental backbones with score-threshold 0.50, with AP75. We have selected top three ResNet architectures with 78.7\%, 77.5\% and 78.8\% mean average precision (mAP) respectively and SSD$\_$MobileNet$\_$v1 with 92.5\% average precision (AP) for line$\_$collapse defect as our final candidate backbones for proposed ensemble model framework while discarding the others for poor average precision accuracy per defect class. For the RetinaNet model, focal loss strategy is implemented with weighting factor of $\alpha=0.25$ and focusing parameter $\gamma=2.0$, to tackle the class imbalance problem as well as to learn from challenging defect instances. Table~\ref{tab:TEST-VALIDATION} presents Test and Validation detection accuracies of top 3 ResNet architecture backbones per defect class along with average inference time in seconds. The proposed RetinaNet framework is an ensemble architecture based on a selective permutation of backbones as ResNet50-to-ResNet152-to-ResNet101 Ref.~\citenum{Dey22}. The selection criterion as described in 4.2, is justified to propose a preference-based ensemble strategy to combine the output predictions from different models and achieve better performance on classification and detection of different defects. As presented in Table~\ref{tab:OVERAL}, our proposed ensemble approach achieves better results with overall mAP of 81.6\% than the results obtained by 3 top individual backbones separately as shown in Table~\ref{tab:TEST-VALIDATION}. There is further scope of improvement of overall mAP metric again considering ensembling of SSD$\_$MobileNet$\_$v1 architecture as a backbone with 92.5\% average precision (AP) for line$\_$collapse defect as a future work.

\begin{table*}[ht]
	\caption{IMPLEMENTATION RESULTS WHEN EXPERIMENTING WITH DIFFERENT BACKBONE ARCHITECTURES. IOU @ 0.50/AP50} 
	\label{tab:IMPLEMENTATION}
	\begin{center}       
		\begin{tabular}{|l|l|l|l|l|l|l|l|} 
			\hline
			\rule[-1ex]{0pt}{3.5ex}  Class Name  &	\footnotesize{ResNet50}  &	\footnotesize{ResNet101}  &	\footnotesize{ResNet152}  & \footnotesize{ MobileNet224$\_$1.0}  &	\footnotesize{SeResNet34}	 & \footnotesize{Vgg19} & 	\footnotesize{Vgg16}   \\
			\hline
			
			\rule[-1ex]{0pt}{3.5ex}  gap$\_$AP &	0.954	& \textbf{0.968}	& 0.963	& 0.462	& 0.034	& 0.958	& 0.933  \\
			
			\hline
			\rule[-1ex]{0pt}{3.5ex} p$\_$gap$\_$AP	& \textbf{0.432}	&0.291	&0.376	&0.00	&0.00	&0.118	&0.235 \\
			
			\hline
			\rule[-1ex]{0pt}{3.5ex} bridge$\_$AP	&\textbf{0.872}	&0.811	&0.844	&0.723	&0.717	&0.732	&0.786\\
			
			\hline
			\rule[-1ex]{0pt}{3.5ex}  microbridge$\_$AP	&0.603	&0.633	&0.669	&0.104	&0.003	& 0.7	& \textbf{0.715}\\
			
			\hline
			\rule[-1ex]{0pt}{3.5ex} line$\_$collapse$\_$AP	& 0.828 	&0.816	&0.789	&\textbf{0.925}	&0.925	&0.799	&0.788\\
			
			\hline
			\rule[-1ex]{0pt}{3.5ex}  \textbf{mAP}	&\textbf{0.787} &	\textbf{0.775}	&\textbf{0.788}	&0.429	&0.222	&0.754	&0.762\\
			
			\hline

			\hline

		\end{tabular}
	\end{center}
\end{table*} 


\begin{table*}[ht]
	\caption{IMPLEMENTATION RESULTS WHEN EXPERIMENTING WITH DIFFERENT BACKBONE ARCHITECTURES. IOU @ 0.75/AP75} 
	\footnotesize
	\label{tab:OVERAL75}
	\begin{center}       
		\begin{tabular}{|l|l|l|l|l|l|l|l|} 
			\hline
			
			
			\hline
			\rule[-1ex]{0pt}{3.5ex}	Class Name	& Resnet50	& Resnet101	& Resnet152 & Mobilenet224$\_$1.0	& SeResnet34	& Vgg19	& vgg16 \\
			
			\hline
			
			\rule[-1ex]{0pt}{3.5ex}	gap$\_$AP	& 0.891	& 0.898	& \textbf{0.929}	& 0.366	& 0.034	& 0.892	& 0.713 \\
			
			\hline
			\rule[-1ex]{0pt}{3.5ex}	p$\_$gap$\_$AP	& \textbf{0.374}	& 0.269	& 0.261	& 0	& 0	& 0.065	& 0.127 \\
			\hline
			\rule[-1ex]{0pt}{3.5ex}	bridge$\_$AP	& 0.434	& \textbf{0.568}	& 0.289	& 0.379	& 0.505	& 0.343	& 0.46 \\
			
			\hline
			
			\rule[-1ex]{0pt}{3.5ex}		microbridge$\_$AP	& 0.603	& 0.611	& \textbf{0.634}	& 0.07	& 0	& \textbf{0.654}	& 0.63 \\
			
			\hline
			
			\rule[-1ex]{0pt}{3.5ex}	line$\_$collapse$\_$AP	& 0.694	& \textbf{0.816}	& 0.789	& 0.232	&0.565	& 0.799	& 0.781 \\

			\hline
			\rule[-1ex]{0pt}{3.5ex}	mAP	& 0.707 &	\textbf{0.727}	& \textbf{0.727}	& 0.234 &	0.144	& 0.692	& 0.619 \\
			
			\hline
			
		\end{tabular}
	\end{center}
\end{table*} 



\begin{table*}[ht]
	\caption{TEST/VALIDATION ACCURACY OF TOP 3 RESNET ARCHITECTURE BACKBONES.} 
	\label{tab:TEST-VALIDATION}
	\begin{center}       
		\begin{tabular}{|l|l|l|l|l|l|l|} 
			\hline
			\rule[-1ex]{0pt}{3.5ex}    Score$\_$threshold: 0.5	& \multicolumn{3}{c|}{Test} &	\multicolumn{3}{c|}{Validation} \\
			\hline
			
			\rule[-1ex]{0pt}{3.5ex} Class Name	& ResNet50 &	ResNet101	& ResNet152	& ResNet50 &	ResNet101	& ResNet152\\
			\hline
			
			\rule[-1ex]{0pt}{3.5ex} gap$\_$AP	&0.954	& \textbf{0.968}	&0.963	& \textbf{0.969}	&0.963	&0.947\\
			\hline
			
			\rule[-1ex]{0pt}{3.5ex} p$\_$gap$\_$AP	& \textbf{0.432}	&0.291	&0.376	& \textbf{0.346}	&0.232	&0.28\\
			\hline
			
			\rule[-1ex]{0pt}{3.5ex} bridge$\_$AP	& \textbf{0.872}	&0.811	&0.844	&0.927	&0.894	& \textbf{0.947}\\
			\hline
			
			\rule[-1ex]{0pt}{3.5ex} microbridge$\_$AP	&0.603	&0.633	& \textbf{0.669}	&0.738	& \textbf{0.792}	&0.786\\
			\hline
			
			\rule[-1ex]{0pt}{3.5ex} line$\_$collapse$\_$AP	& \textbf{0.828}	&0.816	&0.789	& \textbf{0.909}	&0.864	&0.864\\
			\hline
			
			\rule[-1ex]{0pt}{3.5ex} mAP	& 0.787	&0.775	& \textbf{0.788}	& \textbf{0.832}	&0.809	&0.811 \\
			\hline
			
			\rule[-1ex]{0pt}{3ex} average inference 	&0.0769&	\textbf{0.0656}&	0.0782	&--	&--&	--\\
			
			\rule[-1ex]{0pt}{3ex}  time (Secs)	&&	&		&&&	\\
			
			\hline

		\end{tabular}
	\end{center}
\end{table*}


\begin{table}[ht]
	\caption{OVERALL TEST ACCURACY OF PROPOSED RETINANET [ENSEMBLE$\_$RESNET] FRAMEWORK.} 
	\label{tab:OVERAL}
	\begin{center}       
		\begin{tabular}{|l|l|} 
			\hline
			\rule[-1ex]{0pt}{3.5ex}  Proposed Model	& Ensemble$\_$ResNet \\
			\rule[-1ex]{0pt}{3.5ex} & 	[ResNet50$\rightarrow$ResNet152$\rightarrow$ResNet101]   \\
			\hline

			\rule[-1ex]{0pt}{3.5ex} gap$\_$AP	&0.959\\
			\hline
			
			\rule[-1ex]{0pt}{3.5ex} p$\_$gap$\_$AP&	0.52\\
			\hline
			
			\rule[-1ex]{0pt}{3.5ex} bridge$\_$AP &	0.867\\
			\hline
			
			\rule[-1ex]{0pt}{3.5ex} microbridge$\_$AP &	0.675\\
			\hline
			
			\rule[-1ex]{0pt}{3.5ex} line$\_$collapse$\_$AP &	0.828\\
			\hline
			
			\rule[-1ex]{0pt}{3.5ex} mAP	 &\textbf{0.816}\\
			
			\hline

			\hline

		\end{tabular}
	\end{center}
\end{table} 


\begin{figure}
	\begin{center}
		\begin{tabular}{c}
			\includegraphics[width=.700\linewidth]{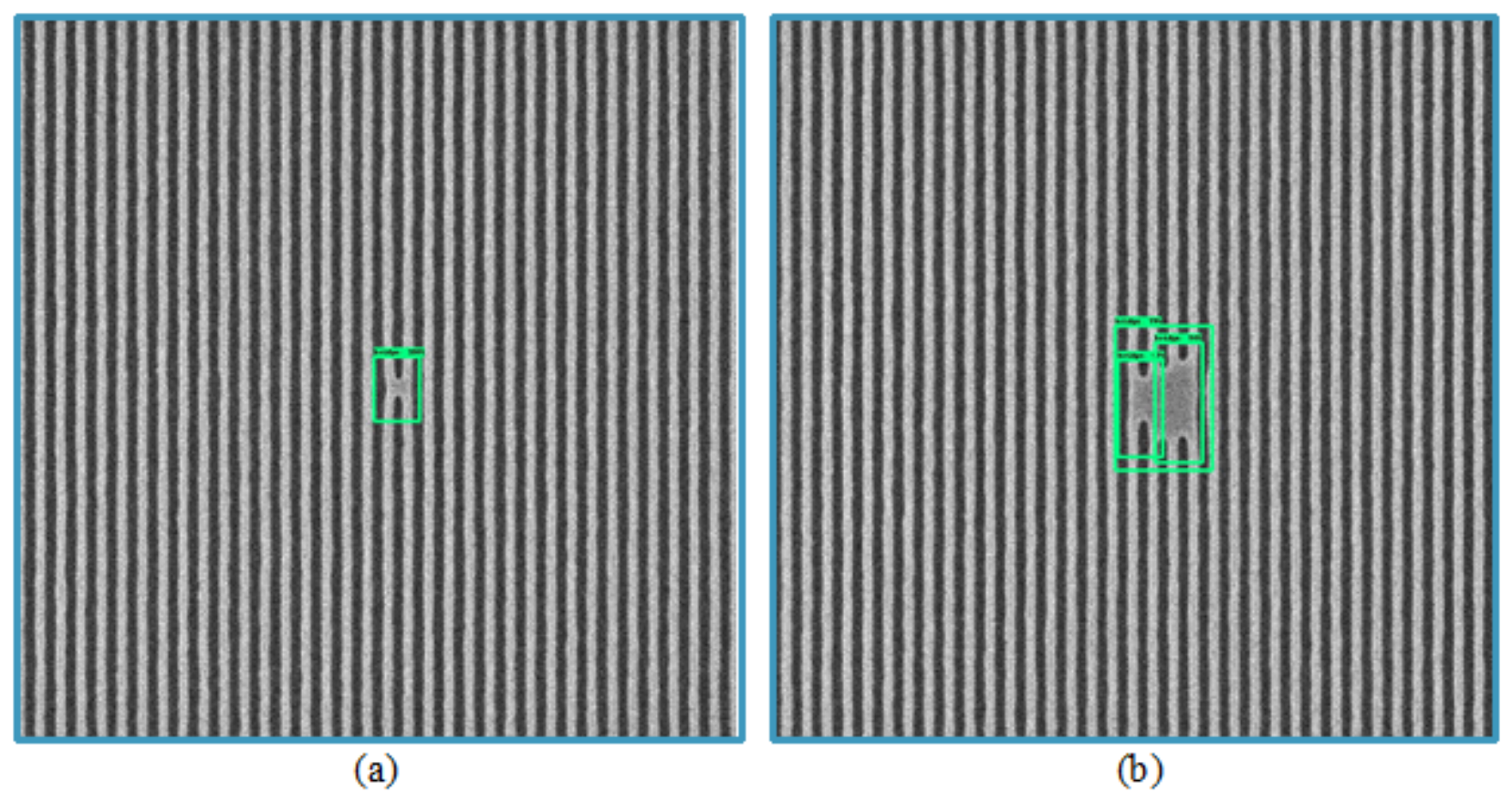}
		\end{tabular}
	\end{center}
	\caption 
	{ \label{fig:BRIDGE}
		BRIDGE detection results with confidence score.
		(a) Single Bridge, (b) Multiple Bridges. } 
\end{figure}

\begin{figure}
	\begin{center}
		\begin{tabular}{c}
			\includegraphics[width=.700\linewidth]{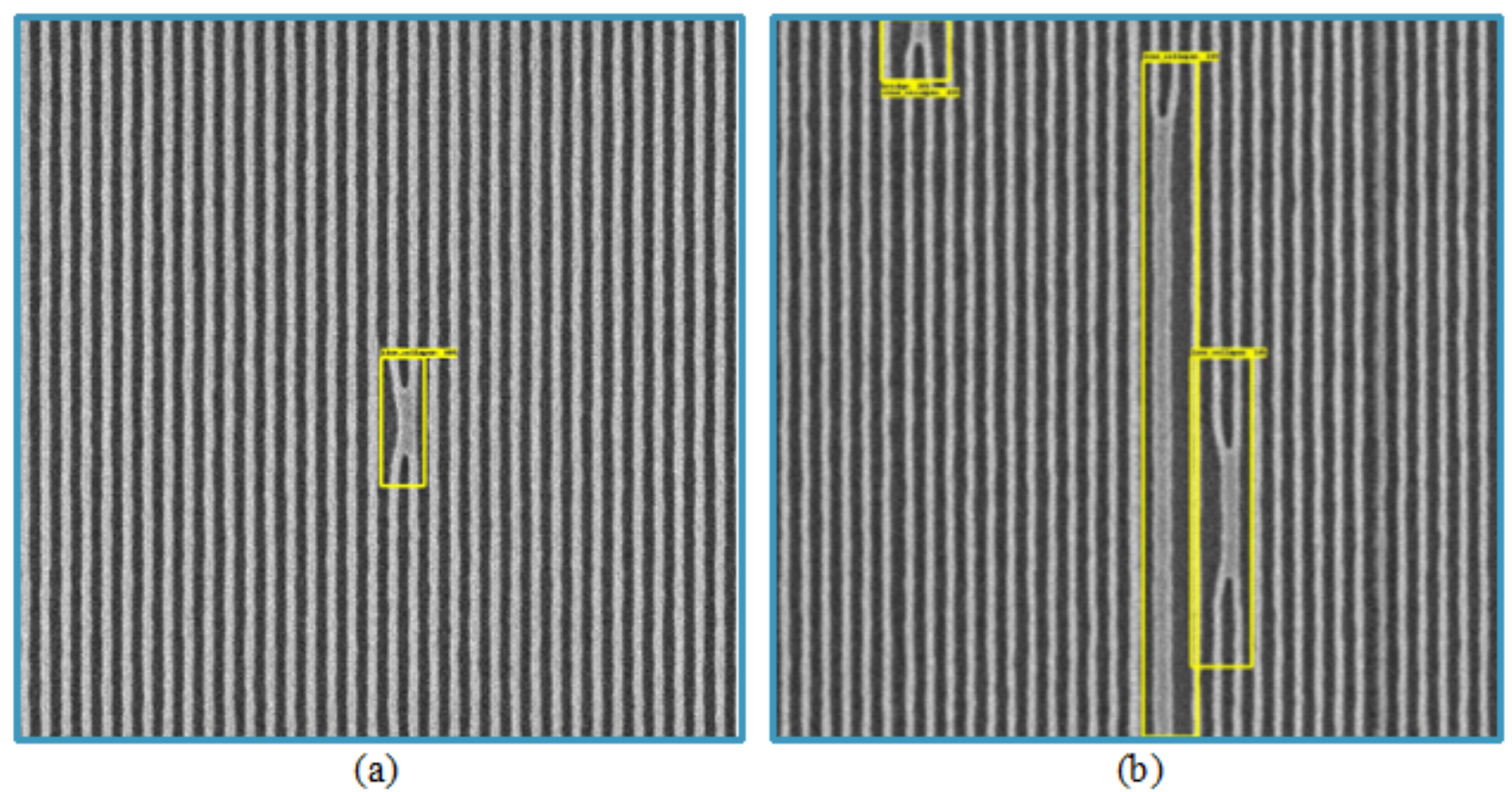}
		\end{tabular}
	\end{center}
	\caption 
	{ \label{fig:LINE-COLLAPSE}
		LINE-COLLAPSE detection results with confidence score. (a) Single Line-Collapse, (b) Multiple Line-Collapses. } 
\end{figure}

\section{Evaluation}
\label{Evaluation}

\subsection{Defect Detection Performance:}

\begin{figure}
	\begin{center}
		\begin{tabular}{c}
			\includegraphics[width=.400\linewidth]{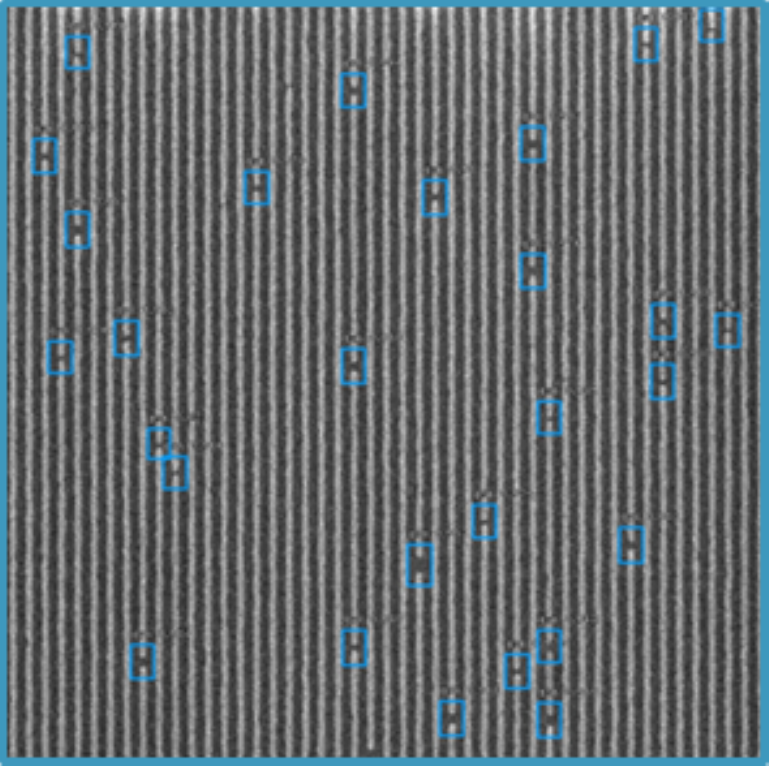}
		\end{tabular}
	\end{center}
	\caption 
	{ \label{fig:GAP}
		GAP/BREAK detection results with confidence score. Multiple Line-breaks. } 
\end{figure}

\begin{figure}
	\begin{center}
		\begin{tabular}{c}
			\includegraphics[width=.700\linewidth]{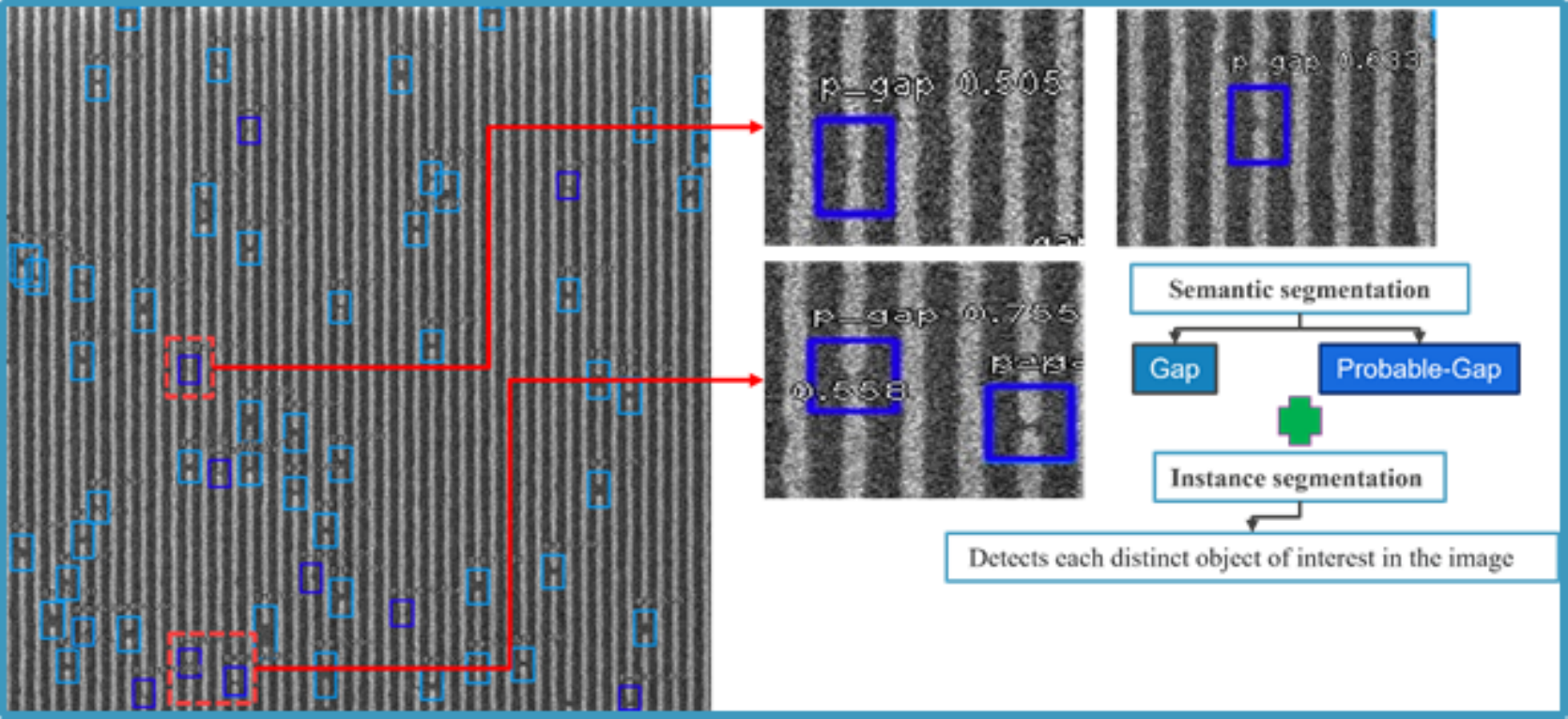}
		\end{tabular}
	\end{center}
	\caption 
	{ \label{fig:Detection}
		Detection results of more challenging Probable NANO-GAP separately in presence
		of  GAP/BREAK. Model shows robustness in detecting relatively few probable
		Nano-Gap defects in presence of  frequent Gap defectivity. } 
\end{figure}

\begin{figure}
	\begin{center}
		\begin{tabular}{c}
			\includegraphics[width=.700\linewidth]{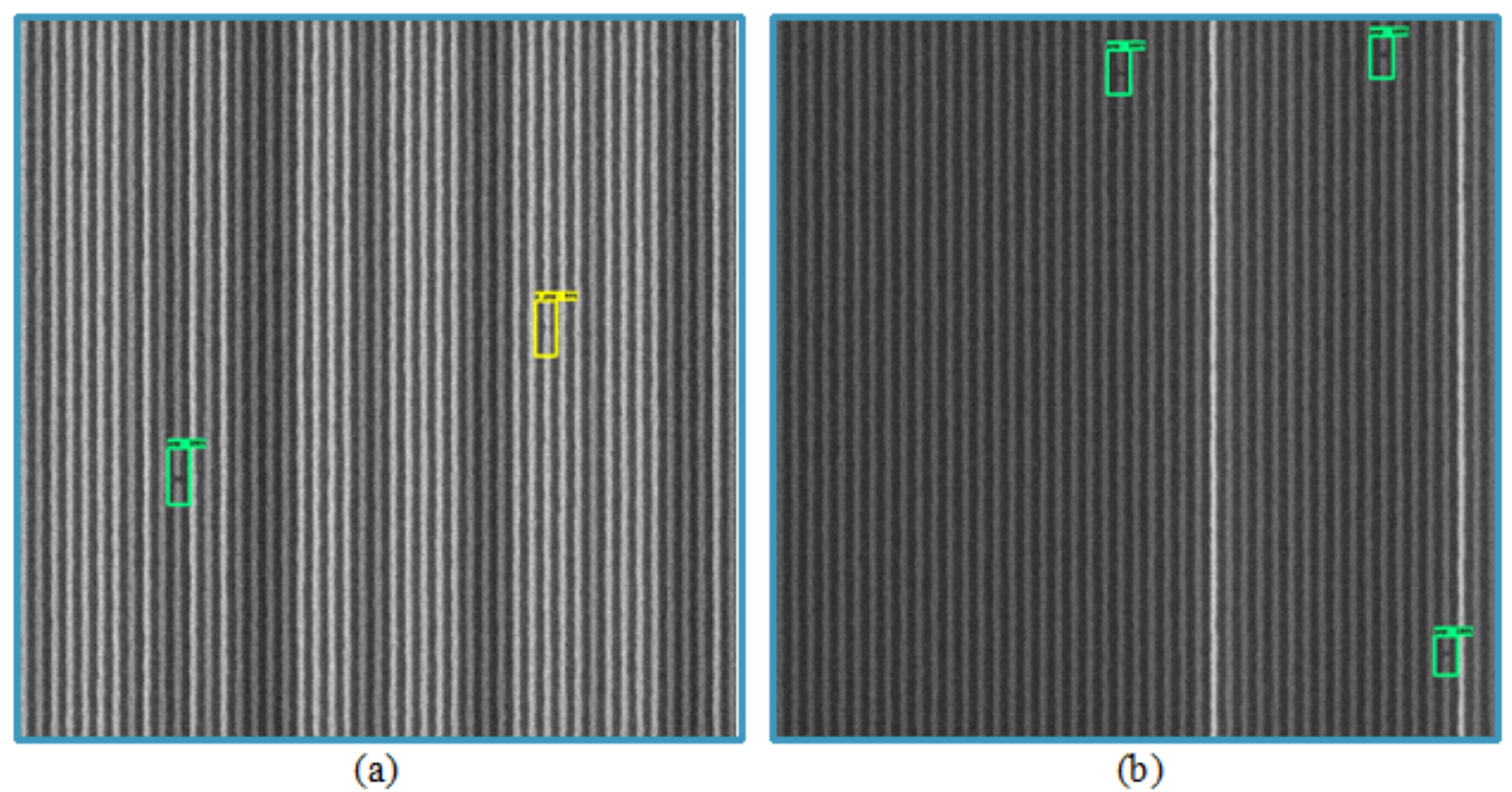}
		\end{tabular}
	\end{center}
	\caption 
	{ \label{fig:NANO-GAP}
		Detection of NANO-GAPs and Probable NANO-GAPs in presence of
		contrast change scenario. Contrast change does not affect defect detection
		performance of proposed ML model in comparison to conventional
		approach.  } 
\end{figure}

\begin{figure}
	\begin{center}
		\begin{tabular}{c}
			\includegraphics[width=.400\linewidth]{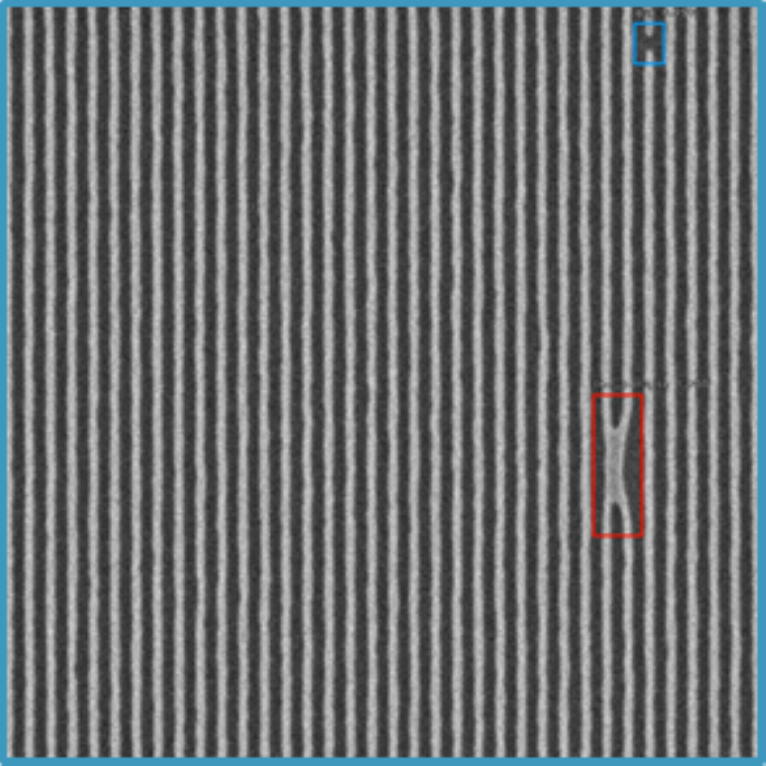}
		\end{tabular}
	\end{center}
	\caption 
	{ \label{fig:Mixed}
		Mixed defect detection results with confidence score.
		(a) Single Line-Collapse, (b) Single Gap/Break.} 
\end{figure}

\begin{figure}
	\begin{center}
		\begin{tabular}{c}
			\includegraphics[width=.600\linewidth]{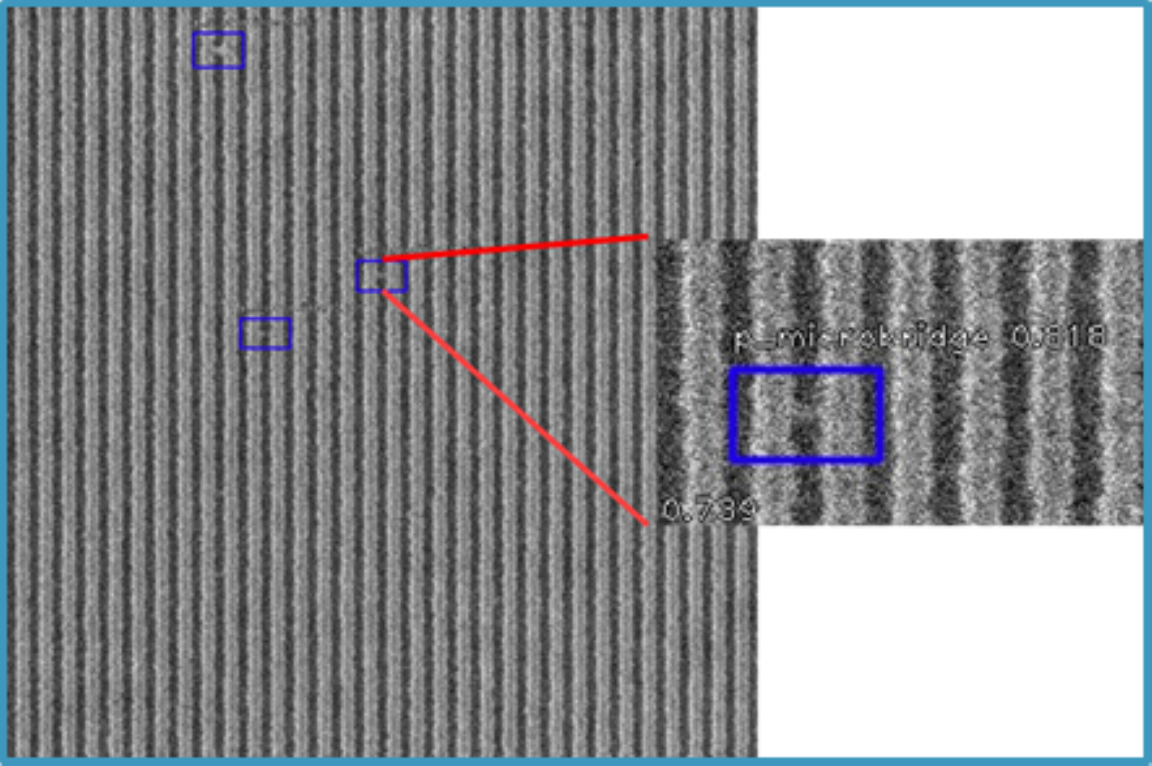}
		\end{tabular}
	\end{center}
	\caption 
	{ \label{fig:NANO-BRIDGE}
		Detection results of more challenging NANO-BRIDGE/MICRO-BRIDGE
		defectivity. } 
\end{figure}

\begin{figure}
	\begin{center}
		\begin{tabular}{c}
			\includegraphics[width=.700\linewidth]{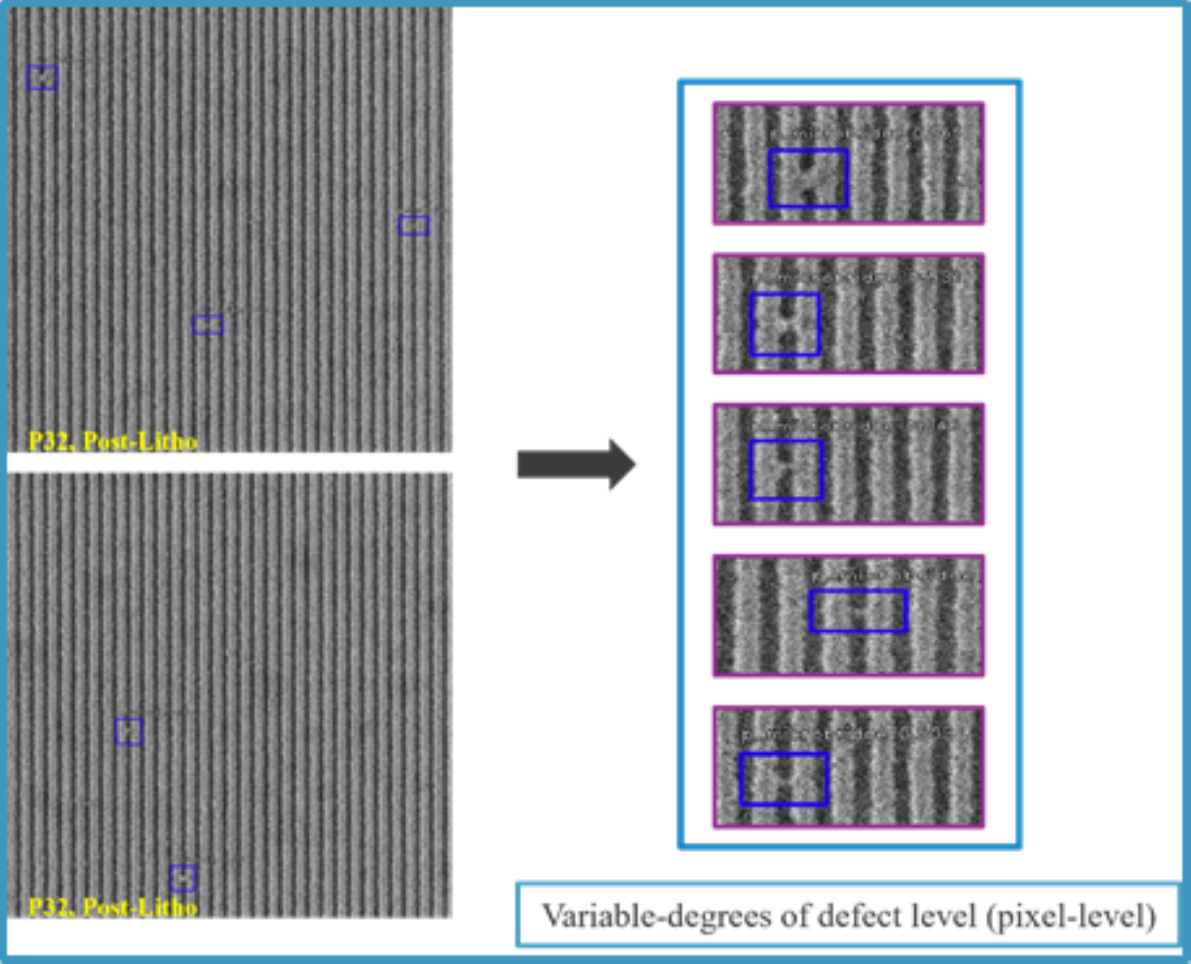}
		\end{tabular}
	\end{center}
	\caption 
	{ \label{fig:NANO-BRIDGE9}
		Detection results of more challenging NANO-BRIDGE/MICRO-BRIDGE
		defectivity on new TEST dataset. Model demonstrates robustness in detecting
		variable degrees of pixel-level micro-bridge defectivity. } 
\end{figure}

\begin{figure}
	\begin{center}
		\begin{tabular}{c}
			\includegraphics[width=.700\linewidth]{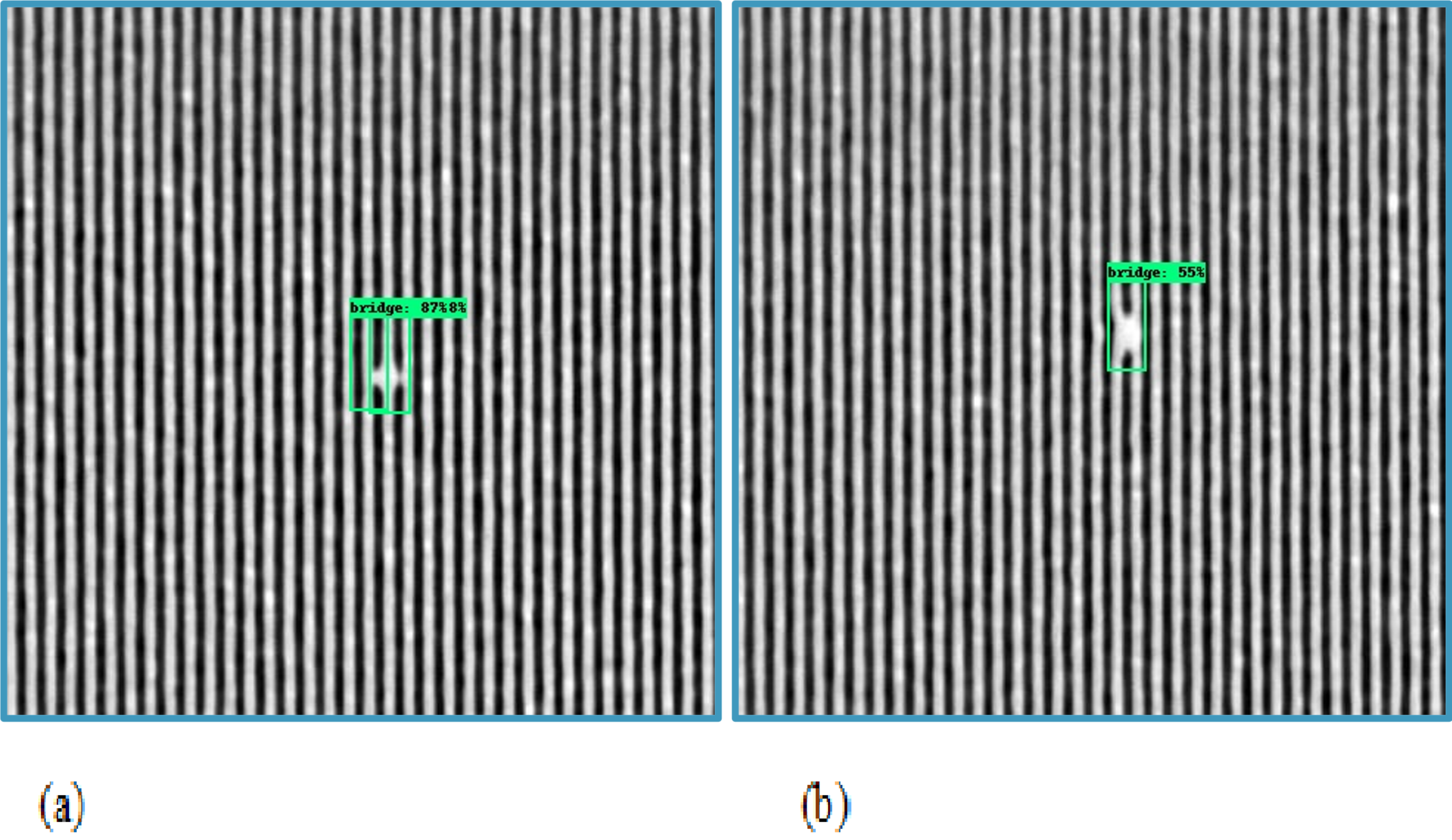}
		\end{tabular}
	\end{center}
	\caption 
	{ \label{fig:jm3_191200}
		Defect detection on Review-SEM images. } 
\end{figure}

Examples of typical defect classification and detection results are shown in [Fig.~\ref{fig:BRIDGE}-Fig.~\ref{fig:GAP}](a) (b) as single or multiple defect instances of bridge, line-collapse and gap/line-break. Fig.~\ref{fig:Detection} shows robustness of our proposed model in detecting relatively few more challenging probable nano-gap defects in presence of  frequent gap defectivity. It emereged as a very challegning scenario for conventional approaches or tools Ref.~\citenum{10.1117/1.JMM.17.4.041011} to differentiate between these two marginal defect categories. However, our proposed model demonstrates (1) semantic segmentation between two distinct defect classes as (a) gap/line-break and (b) probable nano-gap and (2) instance segmentation as detection of each distinct defect of interest under these two defect classes in the same image. Fig.~\ref{fig:NANO-GAP} (a)(b) illustrates detection of nano-gaps and probable nano-gaps in presence of contrast change scenario. We can see contrast change does not affect defect detection performance of our proposed model against conventional approach Ref.~\citenum{10.1117/1.JMM.17.4.041011}. Fig.~\ref{fig:Mixed} depicts defect detection perfromance when mixed defect categories are present in a same image. Fig.~\ref{fig:NANO-BRIDGE}, Fig.~\ref{fig:NANO-BRIDGE9} shows detection results of more challenging nano-bridge/micro-bridge defectivity on new test image dataset. To validate the proposed model performance and robustness, we have run the defect detection inference model on previously unseen SEM image dataset with different resist family. Compostion of a resist is a significant variable that have an impact on the number of stochastic defects as well as its pixelsize like microbridge and probable nano-gap defects, respectively. Our proposed deep learning-based model demonstrates robustness in detecting variable degrees of pixel-level micro-bridge defectivity (detect individual microbridges regardless their extent). Fig.~\ref{fig:jm3_191200} shows defect detection on review-SEM images. Proposed model, as trained with CD-SEM images, shows robust defect detection capability on Review-SEM images (thus different test distribution). Hence, we demonstrated the ability of our proposed framework to generalize over different SEM applications as well as an assist tool for better defect inspection in the production lines of semiconductor industry. Our proposed ensemble model-based defect detection framework achieves the detection precision (AP) of 95.9\% for gap, 86.7\% for bridge, 82.8\% for line$\_$collapse, 67.5\% for microbridge, and 52.0\% for probable nano-gap defectivity, respectively. However, we believe there is further scope of improvement for average precision for specific classes like microbridge and probable nano-gap, thus overall mAP of the proposed framework can also be improved. This will be considered as our next step of this research.

\subsection{Deep Learning Denoiser:}
\begin{figure}
	\begin{center}
		\begin{tabular}{c}
			\includegraphics[width=.800\linewidth]{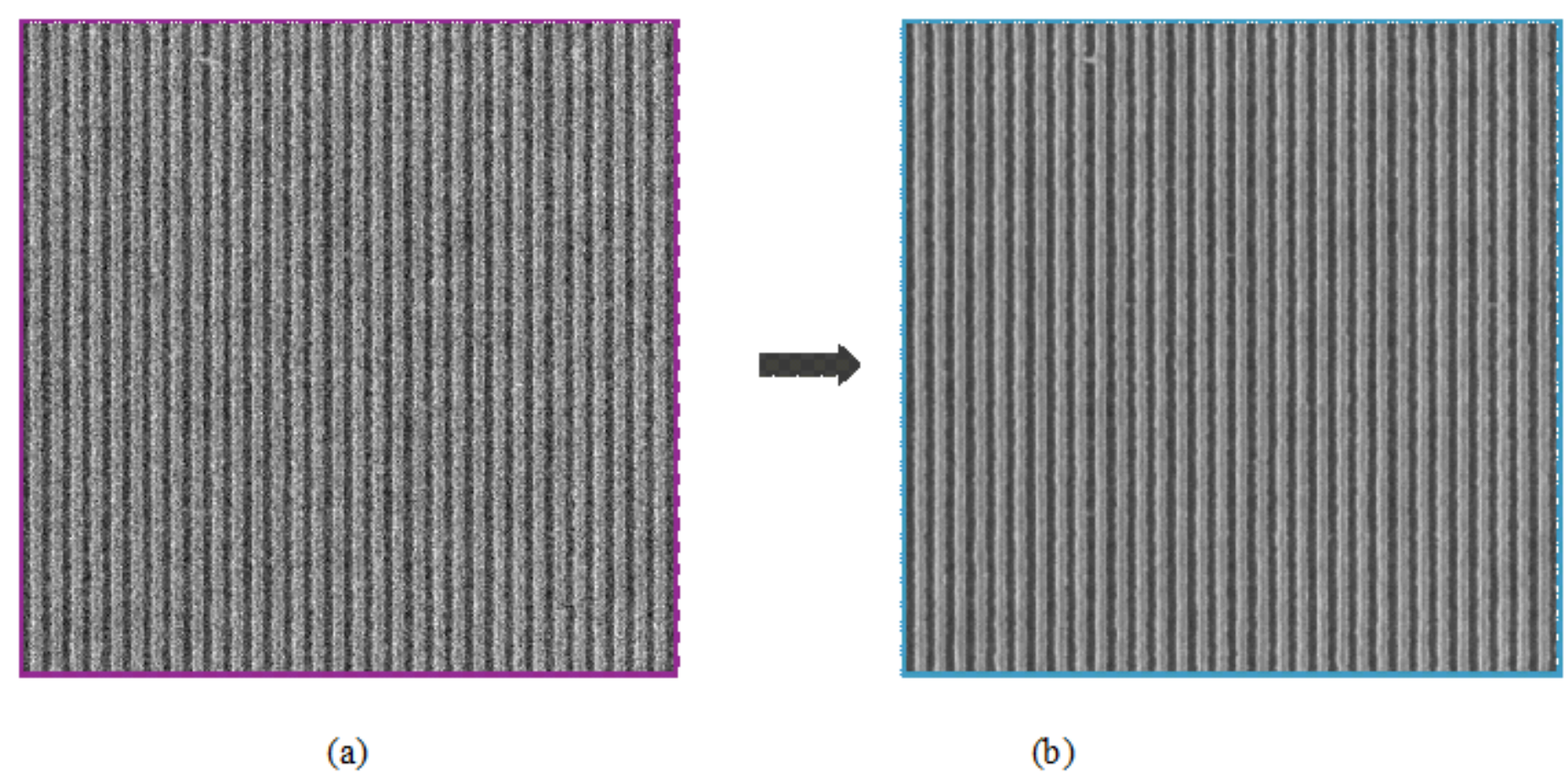}
		\end{tabular}
	\end{center}
	\caption 
	{ \label{fig:Noisy-SEM}
		(a) Noisy SEM image [P32]  with micro/nano-bridges (b) Denoised image. } 
\end{figure}

\begin{figure}
	\begin{center}
		\begin{tabular}{c}
			\includegraphics[width=.800\linewidth]{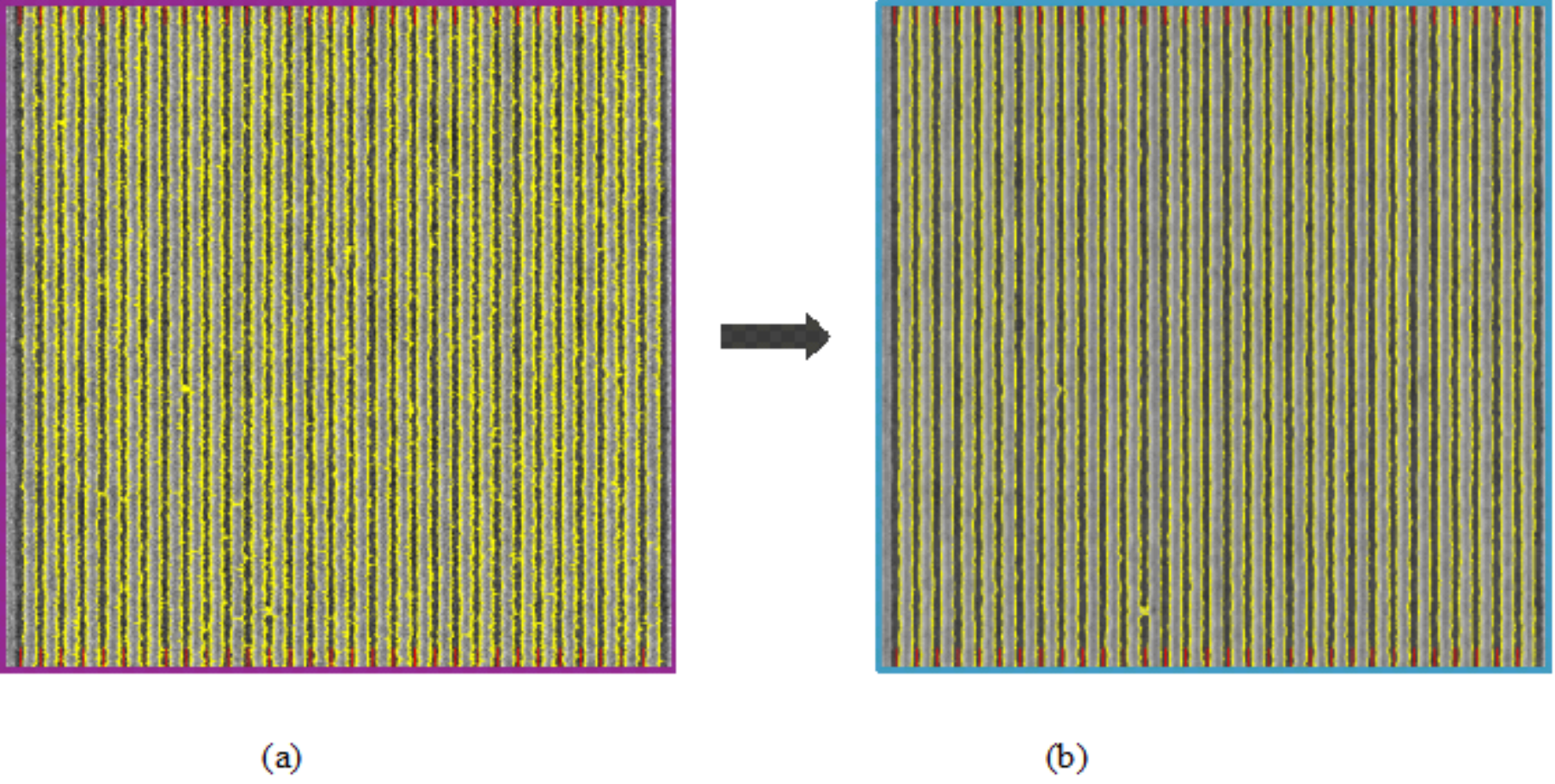}
		\end{tabular}
	\end{center}
	\caption 
	{ \label{fig:SEM}
		SEM image analysis with Fractilia MetroLER library (a) Noisy image (b) Denoised
		image. Detected edges in denoised image are with less spikes or almost without
		spikes in comparison to noisy image. } 
\end{figure}

\begin{figure}
	\begin{center}
		\begin{tabular}{c}
			\includegraphics[width=.800\linewidth]{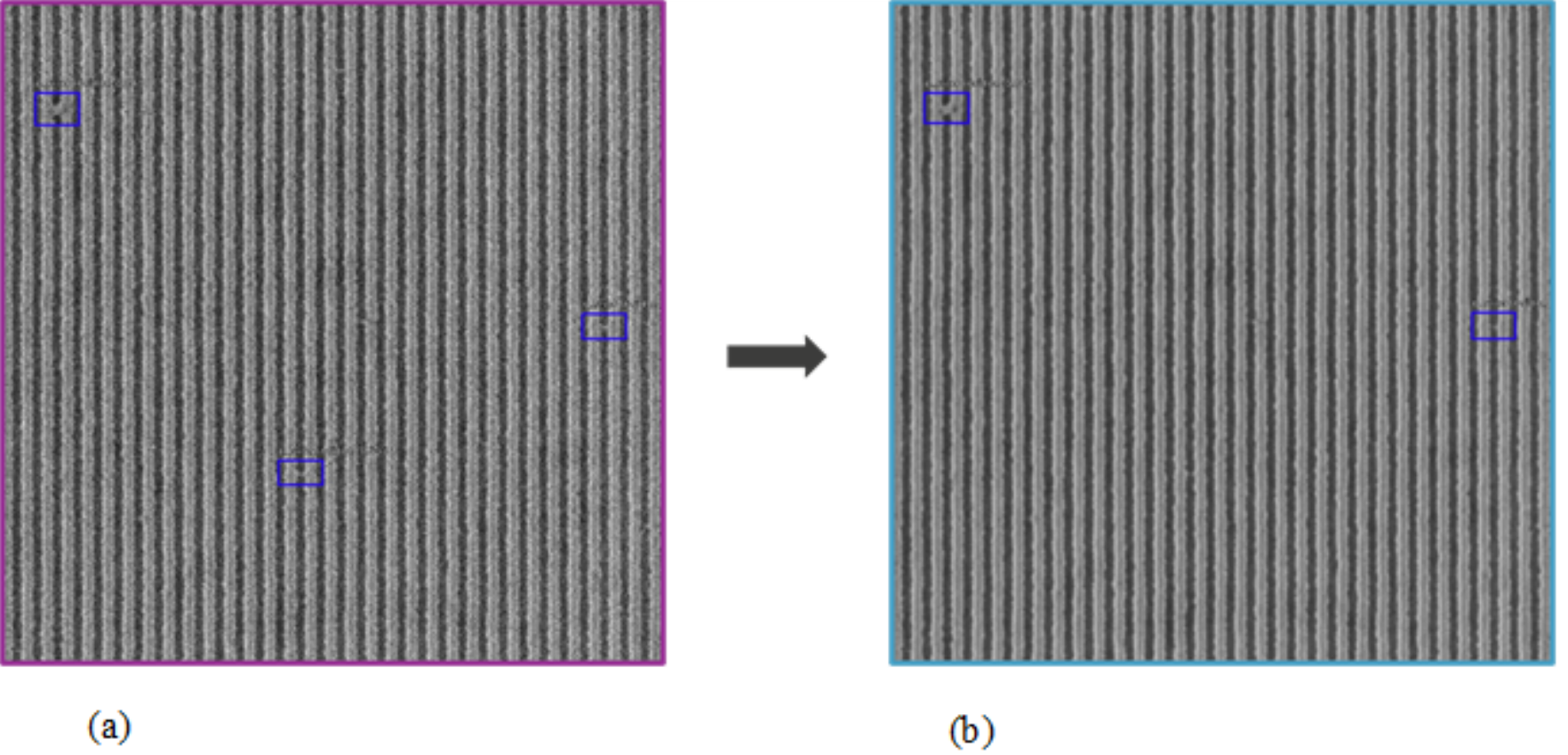}
		\end{tabular}
	\end{center}
	\caption 
	{ \label{fig:Defect-inspection}
		Defect inspection on (a) Noisy SEM image [P32]  with micro/nano-bridges,
		(b) Denoised image. Denoising generally helps to remove FP defects (resist
		footing may appear as micro-bridge defects in presence of stochastic noise on
		structure pixels). } 
\end{figure}

In this section, we have demonstrated how denoising can improve defect inspection performance and accuracy in challenging defect-detection scenarios, specifically in case of micro-bridge detection. The extraction of repeatable and accurate defect locations along with CD metrology becomes significantly complicated in ADI SEM images due to continuous shrinkage of circuit patterns (pitches less than 32 nm). The noise level of SEM images may lead to false defect detections and erroneous metrology. Hence, reducing noise in SEM images is of utmost importance. In Fig.~\ref{fig:Noisy-SEM}, we have shown the denoised SEM image (pitch 32 nm) obtained from the proposed denoising approach Ref.~\citenum{Bap2}. Fig.~\ref{fig:SEM} shows detected edges for denoised image are with less spikes or almost without spikes in comparison to the noisy SEM image when analyzed with Fractilia MetroLER v2.2.5.0 Ref.~\citenum{Chris}. While inspecting a noisy SEM image for microbridge detection, both conventional approach Ref.~\citenum{10.1117/1.JMM.17.4.041011} and our proposed deep learning based approach may flag false positive (FP) defects in terms of resist footing. In presence of stochastic noise on structured pixels, resist footing generally appears as tiny microbridges that are expected to be removed during next etch process step. Denoising optimizes this effect of stochastic noise on structured pixels and therefore, helps to remove the false-positive defects (FP) for better metrology and enhanced defect inspection as demonstrated in Fig.~\ref{fig:Defect-inspection}. We have shown two different strategies in this research as (1) remove any FP detection with strict defect detection confidence score $\geq 0.5$ for microbridge and (2) adaptation of  resist footing as “weak microbridge” defect by lowering enough the confidence score $(0.0 \leqslant score \leqslant 0.5)$. In Fig.~\ref{fig:Defect-detection} and Fig.~\ref{fig:Defect-detection14}, we have presented both approaches. For the first approach, we have repeated the defect inspection step on denoised images with the same trained model parameters with noisy images only, whereas for the later, we have retrained the model with denoised images and fine-tuned the model parameters. Another approach is possible as labeling of resist footing as a new defect category and train the model. This will be considered as our future scope of this research.

\subsection{Comparison with Conventional approach/Tools:}
\begin{figure}
	\begin{center}
		\begin{tabular}{c}
			\includegraphics[width=.800\linewidth]{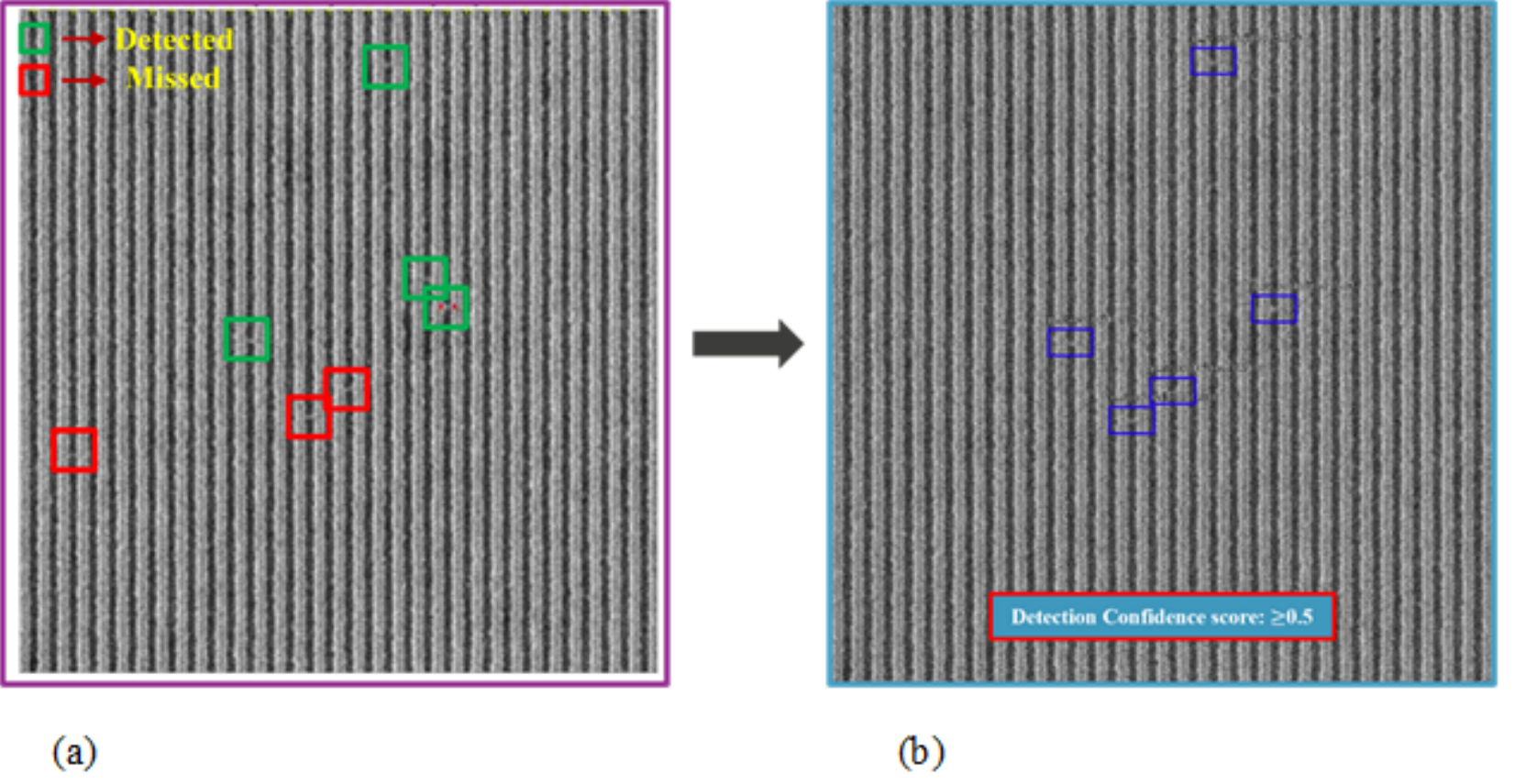}
		\end{tabular}
	\end{center}
	\caption 
	{ \label{fig:Defect-detection}
		Defect detection on same Noisy SEM image [P32]  with micro/nano-bridges:
		(a) Conventional Tool/approach, (b) Proposed ensemble model based approach. } 
\end{figure}

\begin{figure}
	\begin{center}
		\begin{tabular}{c}
			\includegraphics[width=.800\linewidth] {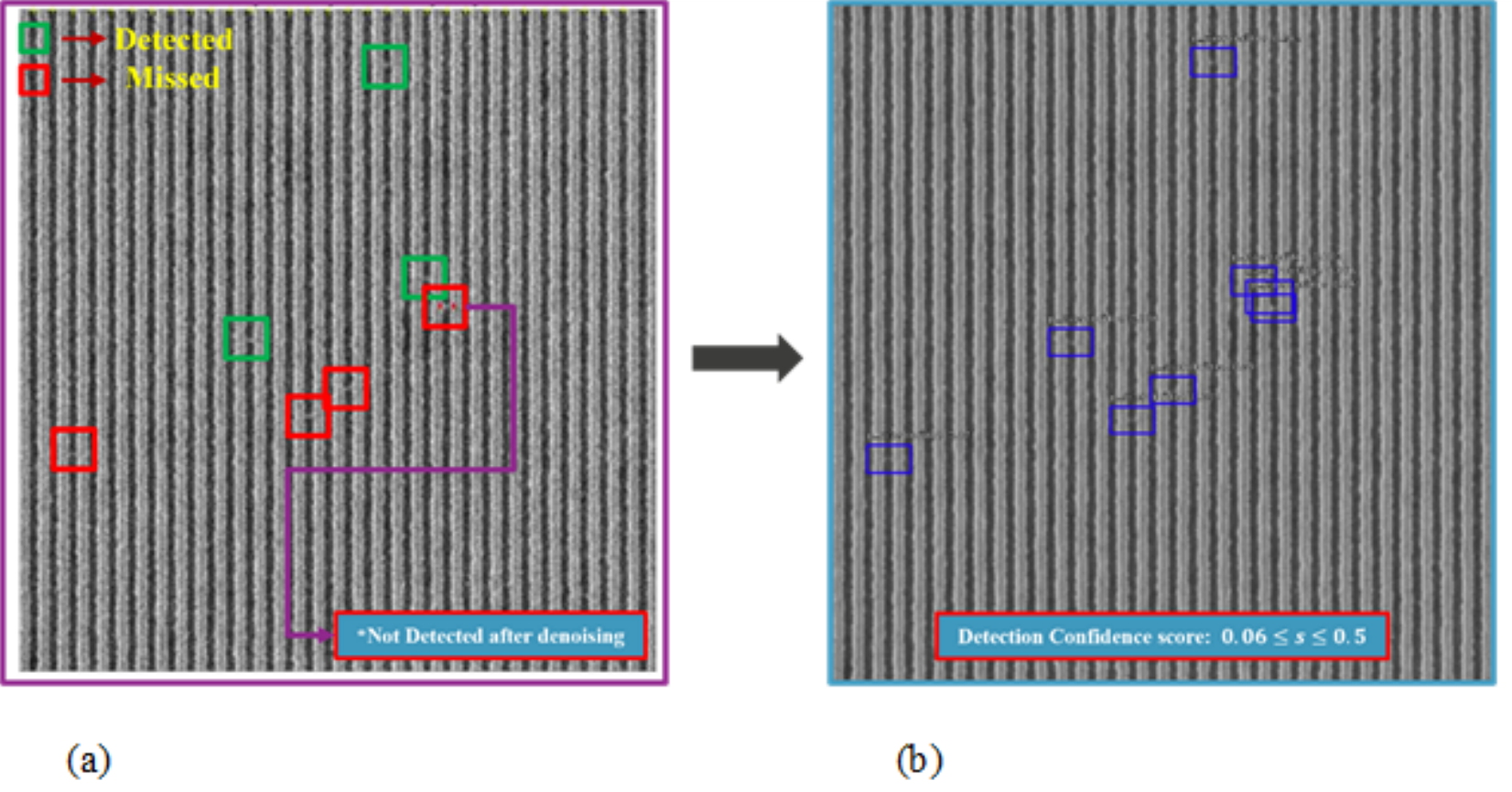}
		\end{tabular}
	\end{center}
	\caption 
	{ \label{fig:Defect-detection14}
		Defect detection on same Denoised SEM image [P32]  with micro/nano-bridges:
		(a) Conventional Tool/approach, (b) Proposed ensemble model based approach. } 
\end{figure}

In this section, we have presented a comparative analysis on stochastic defect detection performance between our proposed deep learning based approach and conventional approach Ref.~\citenum{10.1117/1.JMM.17.4.041011}.  Fig.~\ref{fig:Defect-detection} provides challenging micro/nano-bridges detection scenario on the same Noisy L/S SEM image. With a “manual” selection of the detection threshold parameters (such as user-defined intensity-threshold, failure size parameter, noise etc.), the coventional approach was able to flag four out of seven observable defects. Whereas, our proposed deep learning based model automatically detects five out of the same with a strict defect detection confidence score $\geq 0.5$ without any requirement of such manual trial-and-error based “threshold” selection method. Lowering the automated “confidence score” certainly flags other missing defects as demonstrated in Fig.~\ref{fig:Defect-detection14}. Fig.~\ref{fig:Defect-detection14} demonstrates the same challenging micro/nano-bridges detection scenario on the corresponding denoised image. We can see the detection scenario is influenced by the condition if the image is noisy or denoised for conventional approach. Furthermore, after denoising, along with previous undetected defect instances, the conventional approach was not able to detect the “most obvious” microbridge defect instance which was flagged before. However, our proposed model demonstrates “stable” performance in detecting defects with better accuracy for both noisy or denoised images and replaces the manual trial-and-error based “threshold” selection method with automated “confidence score”. Once defects are correctly detected, different parameters (as length, width, area, additional feature vectors) about the defects can be output for better understanding the root cause of the defects. Thus, Our proposed approach demonstrates its effectiveness both quantitatively and qualitatively.

\begin{figure}
	\begin{center}
		\begin{tabular}{c}
			\includegraphics[width=1.00\linewidth]{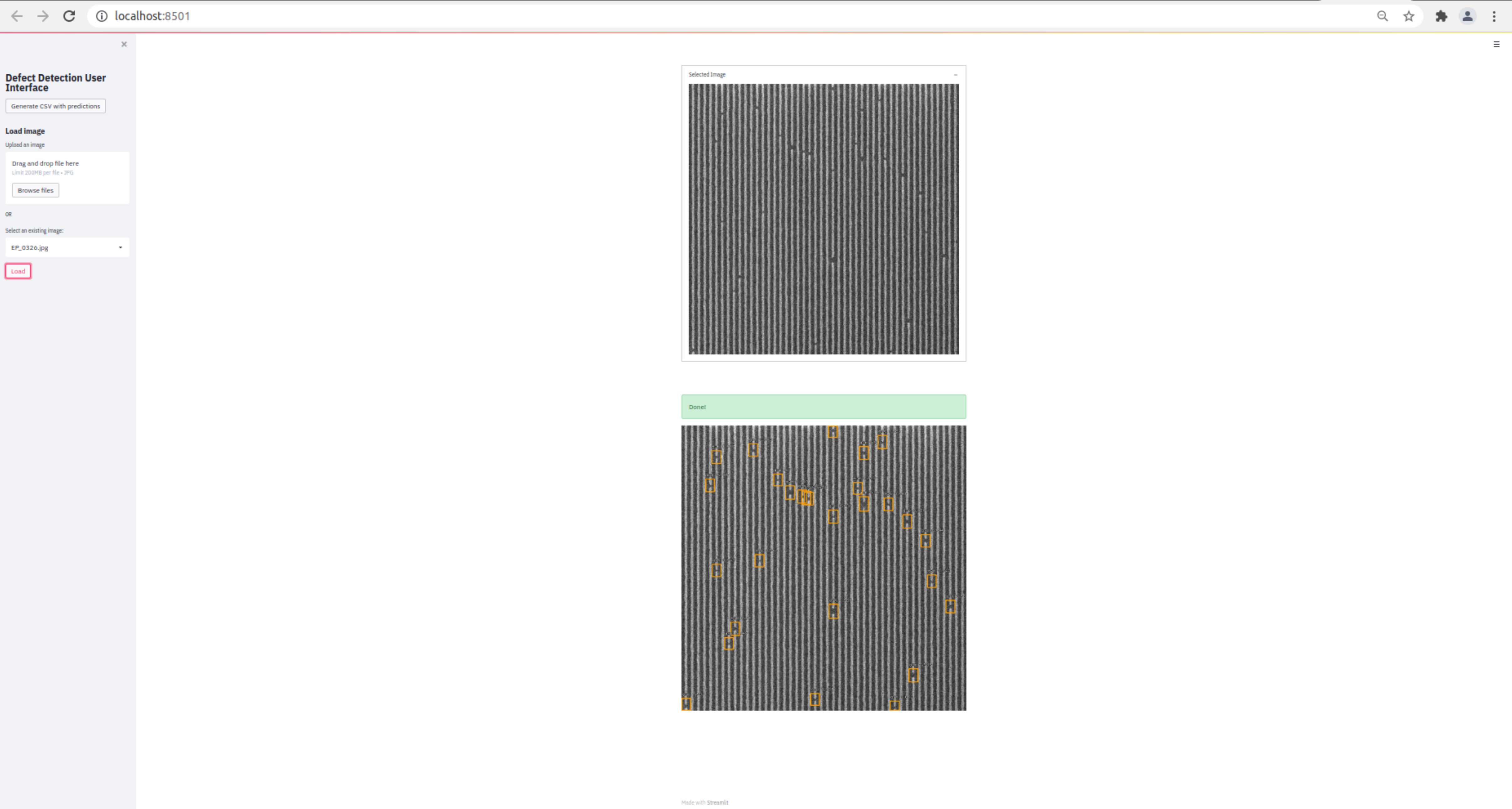}
		\end{tabular}
	\end{center}
	\caption 
	{ \label{fig:Web}
		Web-based defect inspection app. } 
\end{figure} 

\begin{figure}
	\begin{center}
		\begin{tabular}{c}
			\includegraphics[width=0.900\linewidth]{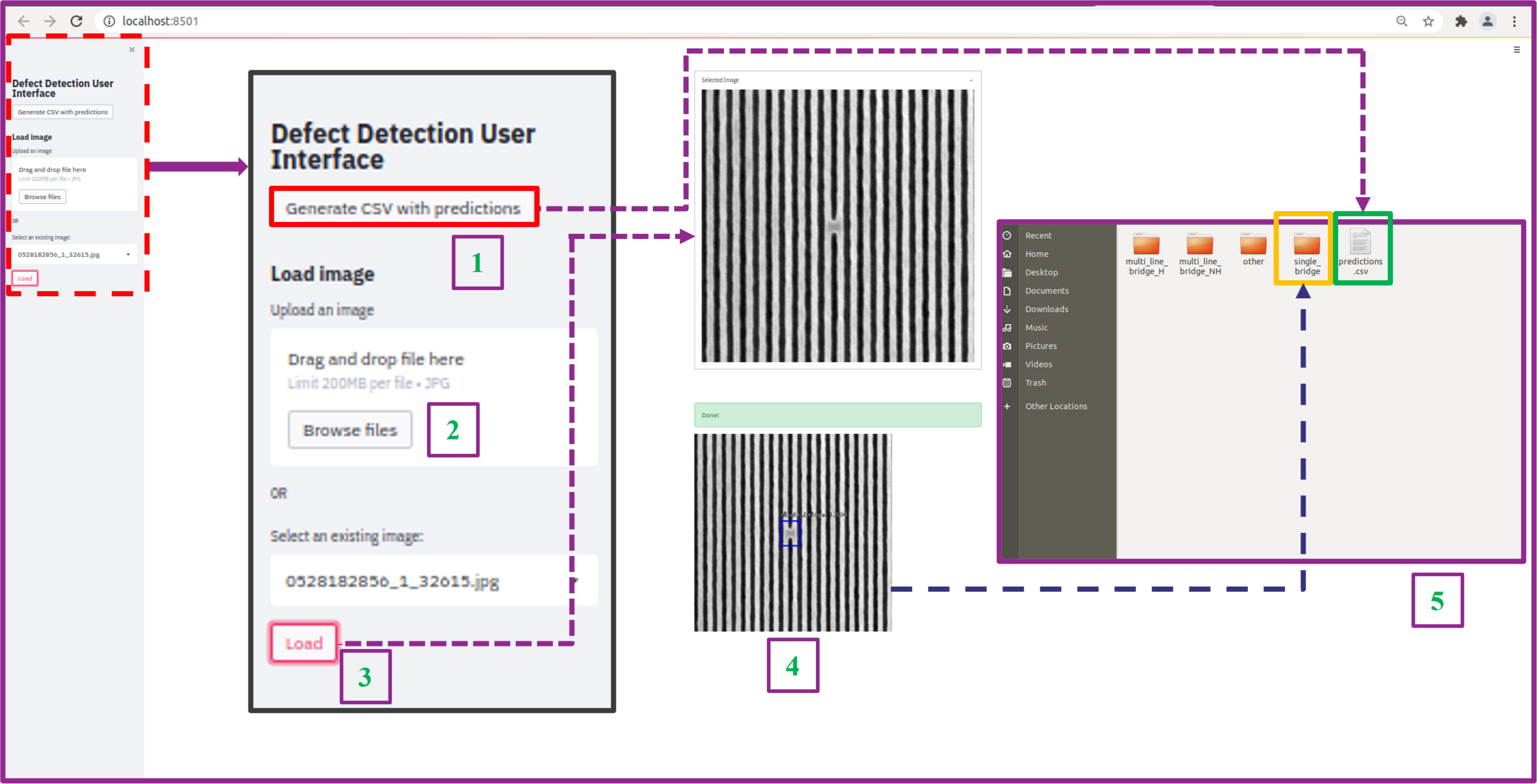}
		\end{tabular}
	\end{center}
	\caption 
	{ \label{fig:Fig-new}
		Essential components of web-based defect inspection app functionality. } 
\end{figure}

\subsection{Defect Classifier User-Interface (UI):}
We built an UI (User-Interface) using Streamlit library in python script to deploy our proposed model as a web-based defect inspection app. A view of the application interface is depicted in Fig.~\ref{fig:Web}. This is a template version of the original proposed software interface and we will add more user-friendly graphical widgets in near future. This will enable different partners/vendors to run the application on their local servers/workstations on their own tool data. This UI will enable the users to upload a dataset of SEM/EDR/Review-SEM images, to select and run one out of different defect detection ineference models on the dataset, to visualize the prediction performance locally and finally to segregate and save the images in different folders according to their defect categorical classes in local machines. As shown in Fig.~\ref{fig:Fig-new}, the essential components of our proposed web-based defect inspection app functionality are as: (1) this graphical widget (“\textit{Generate$\_$CSV$\_$with$\_$predictions}”) enables the users to automatically analyse multiple wafer/DOE data with a single click. Once the defect inspection analysis will be finished, a csv file will be generated containing different parameters (location, area, length, width, corresponding defect class) of the defects for the corresponding wafers and finally to segregate and save the images in different folders according to their defect categorical classes in local machines as shown in (5). In addition, the graphical widget (2) (“\textit{Browse$\_$files}”) allows users to manually browse/load specific wafer data folder of interest, whereas graphical widget (3) (“\textit{Load}”) allows users to manually load and inspect individual image files as shown in (4). Therefore, our proposed web-based defect inspection app helps reduce engineering time as well as tool cycle time against manual inspection method associated with defect inspection process.

\section{Conclusion}
\label{section:Conclusion}
In this work we have developed a novel robust supervised deep learning training scheme to 
accurately classify as well as localize different defect types in SEM images with high degree of accuracy. Our proposed ensemble model is based on a permutation of ResNet50, ResNet101 and ResNet152 architectures as backbones and an experimental selection of preference-based ensemble strategy to combine the output predictions from different models and achieve better performance on classification and detection of defects. Our model demonstrates not only classification of different defect categories as bridges, line-collapses, gaps, micro-bridges, and micro-gaps as well as variable degrees of pixel level defect scenarios for each of these categories. We also have accurately regressed the region-of-defects with a bounding box (represented by center coordinates of box, width, and height) with a detection confidence score. 
Furthermore, we have applied an unsupervised machine learning strategy to denoise the SEM images without the requirement of any clean ground truth or synthetic clean SEM images to remove the False-Positive defects and optimize the effect of stochastic noise on structured pixels. PSD (Power-Spectral-Density) analysis demonstrates that only high frequency component related to noise is affected as expected, keeping the low frequency component, related to the actual morphology of the device feature unaltered. Therefore, our proposed defect inspection pipeline demonstrates enhanced defect detection performance, based on detection accuracy without altering the L/S dimensions in aggressive pitches. As circuit patterns are shrinking to accommodate Moore’s law, the conventional defect inspection procedures are becoming less effective and often leads to false defect detections and erroneous metrology. Our deep learning-based model demonstrates it can overcome the limitations while improving classification, detection, and localization of different defect categories with higher accuracy. Our future strategy is to extend this work towards (1) generate defect classes and locations, (2) generate parameters for each and every defect, (3) use data to model defect transfer from litho to etch, and finally (4) expand to other SEM applications (Logic/CH structures) as well as use other sets of images as TEM/AFM etc. Another possibility is (1) experimentation with (a) different other state-of-the-art deep feature extractor networks as backbones as well as (b) different recent detector frameworks as EfficientNet, EfficientDet etc. to further improve the overall mAP accuracy as well as individual defect class mAP accuracy, (2)addition of new defect categories and experiment with fine tuning of the network parameters to further improve the overall mAP metric.

\bibliography{report} 

\begin{thebibliography}{10}

\bibitem{8373144}
Kim, J., Kim, S., Kwon, N., Kang, H., Kim, Y., and Lee, C., ``Deep learning
  based automatic defect classification in through-silicon via process: Fa:
  Factory automation,'' in [{\em 2018 29th Annual SEMI Advanced Semiconductor
  Manufacturing Conference (ASMC)}{\nolinebreak\hspace{0.1em}]},   35--39
  (2018).

\bibitem{Bap}
Dey, B., Cerbu, D., Khalil, K., Halder, S., Leray, P., Das, S., Sherazi, Y.,
  Bayoumi, M.~A., and Kim, R.~H., ``{Unsupervised machine learning based CD-SEM
  image segregator for OPC and process window estimation},'' in [{\em
  Design-Process-Technology Co-optimization for Manufacturability
  XIV}{\nolinebreak\hspace{0.1em}]},  Yuan, C.-M., ed.,  {\bf 11328},  317 --
  327, International Society for Optics and Photonics, SPIE (2020).

\bibitem{he2015deep}
He, K., Zhang, X., Ren, S., and Sun, J., ``Deep residual learning for image
  recognition,'' (2015).

\bibitem{DBLP:journals/corr/SimonyanZ14a}
Simonyan, K. and Zisserman, A., ``Very deep convolutional networks for
  large-scale image recognition,'' in [{\em 3rd International Conference on
  Learning Representations, {ICLR} 2015, San Diego, CA, USA, May 7-9, 2015,
  Conference Track Proceedings}{\nolinebreak\hspace{0.1em}]},  Bengio, Y. and
  LeCun, Y., eds. (2015).

\bibitem{Hinton504}
Hinton, G.~E. and Salakhutdinov, R.~R., ``Reducing the dimensionality of data
  with neural networks,'' ~{\bf 313}(5786),  504--507 (2006).

\bibitem{10.1007/978-3-319-46448-0_2}
Liu, W., Anguelov, D., Erhan, D., Szegedy, C., Reed, S., Fu, C.-Y., and Berg,
  A.~C., ``Ssd: Single shot multibox detector,'' in [{\em Computer Vision --
  ECCV 2016}{\nolinebreak\hspace{0.1em}]},  Leibe, B., Matas, J., Sebe, N., and
  Welling, M., eds.,  21--37, Springer International Publishing, Cham (2016).

\bibitem{timischl2012statistical}
Timischl, F., Date, M., and Nemoto, S., ``A statistical model of signal--noise
  in scanning electron microscopy,'' {\em Scanning}~{\bf 34}(3),  137--144
  (2012).

\bibitem{Bap2}
Dey, B., Halder, S., Khalil, K., Lorusso, G., Severi, J., Leray, P., and
  Bayoumi, M.~A., ``{SEM image denoising with unsupervised machine learning for
  better defect inspection and metrology},'' in [{\em Metrology, Inspection,
  and Process Control for Semiconductor Manufacturing
  XXXV}{\nolinebreak\hspace{0.1em}]},  Adan, O. and Robinson, J.~C., eds.,
  {\bf 11611},  245 -- 254, International Society for Optics and Photonics,
  SPIE (2021).

\bibitem{17}
Zwitch, R., ``{Streamlit}.'' \url{https://docs.streamlit.io/en/stable/}.

\bibitem{khalil2022designing}
Khalil, K., Eldash, O., Kumar, A., and Bayoumi, M., ``Designing novel aad
  pooling in hardware for a convolutional neural network accelerator,'' {\em
  IEEE Transactions on Very Large Scale Integration (VLSI) Systems}  (2022).

\bibitem{726791}
Lecun, Y., Bottou, L., Bengio, Y., and Haffner, P., ``Gradient-based learning
  applied to document recognition,'' {\em Proceedings of the IEEE}~{\bf
  86}(11),  2278--2324 (1998).

\bibitem{khalil2019economic}
Khalil, K., Eldash, O., Kumar, A., and Bayoumi, M., ``Economic {LSTM} approach
  for recurrent neural networks,'' {\em IEEE Transactions on Circuits and
  Systems II: Express Briefs}~{\bf 66}(11),  1885--1889 (2019).

\bibitem{10.5555/2999134.2999257}
Krizhevsky, A., Sutskever, I., and Hinton, G.~E., ``Imagenet classification
  with deep convolutional neural networks,'' in [{\em Proceedings of the 25th
  International Conference on Neural Information Processing Systems - Volume
  1}{\nolinebreak\hspace{0.1em}]},  {\em NIPS'12},  1097–1105, Curran
  Associates Inc., Red Hook, NY, USA (2012).

\bibitem{7298594}
Szegedy, C., Liu, W., Jia, Y., Sermanet, P., Reed, S., Anguelov, D., Erhan, D.,
  Vanhoucke, V., and Rabinovich, A., ``Going deeper with convolutions,'' in
  [{\em 2015 IEEE Conference on Computer Vision and Pattern Recognition
  (CVPR)}{\nolinebreak\hspace{0.1em}]},   1--9 (2015).

\bibitem{8237586}
Lin, T.-Y., Goyal, P., Girshick, R., He, K., and Dollár, P., ``Focal loss for
  dense object detection,'' in [{\em 2017 IEEE International Conference on
  Computer Vision (ICCV)}{\nolinebreak\hspace{0.1em}]},   2999--3007 (2017).

\bibitem{4777721}
Sharifzadeh, M., Alirezaee, S., Amirfattahi, R., and Sadri, S., ``Detection of
  steel defect using the image processing algorithms,'' in [{\em 2008 IEEE
  International Multitopic Conference}{\nolinebreak\hspace{0.1em}]},   125--127
  (2008).

\bibitem{Kim2017AutomaticDD}
Kim, S. and Oh, I.-S., ``Automatic defect detection from sem images of wafers
  using component tree,'' {\em Journal of Semiconductor Technology and
  Science}~{\bf 17},  86--93 (2017).

\bibitem{6552751}
Bonam, R., Halle, S., Corliss, D., Tien, H.-Y., Wang, F., Fang, W., and Jau,
  J., ``E-beam inspection of euv programmed defect wafers for printability
  analysis,'' in [{\em ASMC 2013 SEMI Advanced Semiconductor Manufacturing
  Conference}{\nolinebreak\hspace{0.1em}]},   310--314 (2013).

\bibitem{8765895}
Wang, J., Yang, Z., Zhang, J., Zhang, Q., and Chien, W.-T.~K., ``Adabalgan: An
  improved generative adversarial network with imbalanced learning for wafer
  defective pattern recognition,'' {\em IEEE Transactions on Semiconductor
  Manufacturing}~{\bf 32}(3),  310--319 (2019).

\bibitem{Lee_2019}
Lee, J.-H. and Lee, J.-H., ``A reliable defect detection method for patterned
  wafer image using convolutional neural networks with the transfer learning,''
  {\em {IOP} Conference Series: Materials Science and Engineering}~{\bf 647},
  012010 (oct 2019).

\bibitem{8944584}
Devika, B. and George, N., ``Convolutional neural network for semiconductor
  wafer defect detection,'' in [{\em 2019 10th International Conference on
  Computing, Communication and Networking Technologies
  (ICCCNT)}{\nolinebreak\hspace{0.1em}]},   1--6 (2019).

\bibitem{app10155340}
Chien, J.-C., Wu, M.-T., and Lee, J.-D., ``Inspection and classification of
  semiconductor wafer surface defects using cnn deep learning networks,'' {\em
  Applied Sciences}~{\bf 10}(15) (2020).

\bibitem{8791815}
Yuan-Fu, Y., ``A deep learning model for identification of defect patterns in
  semiconductor wafer map,'' in [{\em 2019 30th Annual SEMI Advanced
  Semiconductor Manufacturing Conference (ASMC)}{\nolinebreak\hspace{0.1em}]},
   1--6 (2019).

\bibitem{10.1117/1.JMM.19.2.024801}
Patel, D.~V., Bonam, R.~K., and Oberai, A.~A., ``{Deep learning-based
  detection, classification, and localization of defects in semiconductor
  processes},'' {\em Journal of Micro/Nanolithography, MEMS, and MOEMS}~{\bf
  19}(2),  1 -- 15 (2020).

\bibitem{dey2021unsupervised}
Dey, B., Wu, S., Das, S., Khalil, K., Halder, S., Leray, P., Bhamidipati, S.,
  Ahi, K., Pereira, M., Fenger, G., et~al., ``Unsupervised machine learning
  based sem image denoising for robust contour detection,'' in [{\em
  International Conference on Extreme Ultraviolet Lithography
  2021}{\nolinebreak\hspace{0.1em}]},   {\bf 11854},  44--58, SPIE (2021).

\bibitem{lin2017feature}
Lin, T.-Y., Dollár, P., Girshick, R., He, K., Hariharan, B., and Belongie, S.,
  ``Feature pyramid networks for object detection,'' (2017).

\bibitem{CasadoGarcia19}
Casado-García, A. and Heras, J., ``Ensemble methods for object detection,''
  (2019).
\newblock \url{https://github.com/ancasag/ensembleObjectDetection}.

\bibitem{chollet2015keras}
Chollet, F. et~al., ``Keras,'' (2015).

\bibitem{tensorflow2015-whitepaper}
Abadi, M., Agarwal, A., Barham, P., Brevdo, E., Chen, Z., Citro, C., Corrado,
  G.~S., Davis, A., Dean, J., Devin, M., Ghemawat, S., Goodfellow, I., Harp,
  A., Irving, G., Isard, M., Jia, Y., Jozefowicz, R., Kaiser, L., Kudlur, M.,
  Levenberg, J., Man\'{e}, D., Monga, R., Moore, S., Murray, D., Olah, C.,
  Schuster, M., Shlens, J., Steiner, B., Sutskever, I., Talwar, K., Tucker, P.,
  Vanhoucke, V., Vasudevan, V., Vi\'{e}gas, F., Vinyals, O., Warden, P.,
  Wattenberg, M., Wicke, M., Yu, Y., and Zheng, X., ``{TensorFlow}: Large-scale
  machine learning on heterogeneous systems,'' (2015).
\newblock Software available from tensorflow.org.

\bibitem{Ingo24}
Lütkebohle, I., ``{LabelImg}.'' \url{https://github.com/tzutalin/labelImg}
  (2015).

\bibitem{8953982}
Rezatofighi, H., Tsoi, N., Gwak, J., Sadeghian, A., Reid, I., and Savarese, S.,
  ``Generalized intersection over union: A metric and a loss for bounding box
  regression,'' in [{\em 2019 IEEE/CVF Conference on Computer Vision and
  Pattern Recognition (CVPR)}{\nolinebreak\hspace{0.1em}]},   658--666 (2019).

\bibitem{kingma2017adam}
Kingma, D.~P. and Ba, J., ``Adam: A method for stochastic optimization,''
  (2017).

\bibitem{Dey22}
E.~Dehaerne, B.~D. and Halder, S.,
  ``En-sembledeeplearning-baseddefectclassificationand detection in sem
  images.,'' (2021).
\newblock
  \url{https://learnopencv.com/ensemble-deep-learning-based-/defect-classification-and-detection-in-sem-images/}.

\bibitem{10.1117/1.JMM.17.4.041011}
Bisschop, P.~D., ``{Stochastic printing failures in extreme ultraviolet
  lithography},'' {\em Journal of Micro/Nanolithography, MEMS, and MOEMS}~{\bf
  17}(4),  1 -- 23 (2018).

\bibitem{Chris}
Mack, C.~A. and Bunday, B.~D., ``{Analytical linescan model for SEM
  metrology},'' in [{\em Metrology, Inspection, and Process Control for
  Microlithography XXIX}{\nolinebreak\hspace{0.1em}]},  Cain, J.~P. and
  Sanchez, M.~I., eds.,  {\bf 9424},  117 -- 139, International Society for
  Optics and Photonics, SPIE (2015).

\end{thebibliography}
\bibliographystyle{spiebib} 

\end{document}